\documentclass[sigconf]{acmart}

\usepackage{multirow}

\AtBeginDocument{%
  }

\copyrightyear{2026}
\acmYear{2026}
\setcopyright{cc}
\setcctype{by-nc-nd}
\acmConference[CHI '26]{Proceedings of the 2026 CHI Conference on Human Factors in Computing Systems}{April 13--17, 2026}{Barcelona, Spain}
\acmBooktitle{Proceedings of the 2026 CHI Conference on Human Factors in Computing Systems (CHI '26), April 13--17, 2026, Barcelona, Spain}
\acmPrice{}
\acmDOI{10.1145/3772318.3791575}
\acmISBN{979-8-4007-2278-3/2026/04}

\begin{document}

\title{Human-centered Perspectives on a Clinical Decision Support System for Intensive Outpatient Veteran PTSD Care}


\author{Cynthia M Baseman}
\authornote{Both authors contributed equally to this research.}
\email{baseman@gatech.edu}
\orcid{0009-0004-8507-438X}
\author{Myeonghan Ryu}
\authornotemark[1]
\email{mhryu@gatech.edu}
\orcid{0009-0005-2066-5418}
\affiliation{%
  \institution{Georgia Institute of Technology}
  \city{Atlanta}
  \state{Georgia}
  \country{USA}
}

\author{Nathaniel Swinger}
\affiliation{%
  \institution{Georgia Institute of Technology}
  \city{Atlanta}
  \state{Georgia}
  \country{USA}}
\email{nswinger3@gatech.edu}
\orcid{0009-0008-6248-5652}

\author{Kefan Xu}
\affiliation{%
  \institution{Georgia Institute of Technology}
  \city{Atlanta}
  \state{Georgia}
  \country{USA}}
\email{kefanxu@gatech.edu}
\orcid{0000-0002-5492-8061}

\author{Andrew M Sherrill}
\affiliation{%
  \institution{Emory University School of Medicine}
  \city{Atlanta}
  \state{Georgia}
  \country{USA}}
\email{andrew.m.sherrill@emory.edu}
\orcid{0000-0002-7743-745X}

\author{Rosa I Arriaga}
\affiliation{%
  \institution{Georgia Institute of Technology}
  \city{Atlanta}
  \state{Georgia}
  \country{USA}}
\email{arriaga@cc.gatech.edu}
\orcid{0000-0002-8642-7245}

\renewcommand{\shortauthors}{Baseman and Ryu et al.}

\begin{abstract}

Psychotherapy delivery relies on a negotiation between patient self-reports and clinical intuition. Growing evidence for technological support of psychotherapy suggests opportunities to aid the mediation of this tension. To explore this prospect, we designed a prototype of a clinical decision support system (CDSS) for treating veterans with post-traumatic stress disorder in a Prolonged Exposure (PE) therapy intensive outpatient program. We conducted a two-phase interview study to collect perspectives from practicing PE clinicians and former PE patients who are United States veterans. Our analysis distills opportunities for a CDSS (e.g., offering homework review at a glance, aiding patient conceptualization) and larger challenges related to context and deployment (e.g., navigating Veterans Affairs). By reframing our findings through three human-centered perspectives (distributed cognition, situated learning, infrastructural inversion), we highlight the complexities of designing a CDSS for psychotherapists in this context and offer theory-aligned design considerations.

\end{abstract}



\maketitle

\section{Introduction}

Recent literature suggests various technological opportunities to support psychotherapy delivery and bolster patient engagement. Research in human-computer interaction (HCI) and related fields has aimed to support the treatment of mental health conditions including depression \cite{rohani_mubs_2020, rohani_personalizing_2019, kawanishi_lifelog-based_2015, schueller_initial_2015},  insomnia \cite{zhu_making_2017}, and post-traumatic stress disorder (PTSD) \cite{kuhn_randomized_2017, reger2013pe}. Importantly, psychotherapy practice is reliant on a negotiation between patient self-reports and clinical intuition. Recent explorations of sensor-captured patient-generated data (sPGD) collected in the wild suggest that sPGD may empower patients \cite{evansUsingSensorCapturedPatientGenerated2024,ng2019provider,khatiwada2024patient} and facilitate data-driven communication between clinicians and patients \cite{mishra2018supporting, schroeder_supporting_2017} by providing additional context for conversation. It remains an open question, however, what role technology should play in mediating the space between patient self-reports and clinical intuition, and what implications could surface by introducing a non-human intermediary at the site of that tension.

Clinicians make decisions about a patient's treatment informed in part by this negotiation between patient self-report and their own intuition. This suggests an opportunity for a clinical decision support system (CDSS) in making sense of these two potentially divergent sources of information. CDSSs offer opportunities to facilitate the clinical workflow, for example through assisting diagnosis \cite{beede_human-centered_2020, wang_brilliant_2021}, medication prescription \cite{jo_designing_2022, jacobs_designing_2021}, patient evaluation \cite{lee_towards_2022, lee_human-ai_2021} and sense-making \cite{jin2017collaborative}. The importance of taking a human-centered approach to design is underscored by the large body of work on the complexities of user needs and importance of contextual fit for CDSSs \cite{ yang_investigating_2016,  cai_hello_2019, khairat_reasons_2018, wang_human-centered_2023, zhang2022get, jo_designing_2022}. These concerns indicate persisting knowledge gaps in how to design CDSSs for successful integration within real-world clinical contexts. Designing psychotherapist-facing technologies is further complicated by the challenge of gaining a deep understanding of sensitive clinical contexts without posing a risk to patient safety or privacy (i.e., through direct observation). 

This work explores opportunities for a clinician-facing interface to facilitate the delivery of psychotherapy, specifically in the context of United States veterans with post-traumatic stress disorder (PTSD). We focus our attention on an intensive outpatient program, subsidized state-of-the-art care that is offered free of charge to U.S. veterans \cite{WarriorCareNetwork}. This CDSS offers an interesting case study because it is a collaborative clinical setting in which patients and clinicians must efficiently (in just two weeks) work towards a common goal of patient health. It is also a learning context in which a patient learns how to process a traumatic memory. At the same time, the veterans' broader experiences with the United States health care system are heavily mediated by institutional and organizational structures such as the Department of Veterans Affairs (the VA).

As a technology probe, we designed a Figma prototype of a clinician dashboard to support the delivery of Prolonged Exposure (PE) therapy. PE therapy is an evidence-based cognitive behavioral therapy for treating PTSD \cite{bisson2013psychological, rothbaum2002exposure}. PE utilizes experiential exercises to teach patients how to gradually approach reminders of a traumatic memory without avoidance. These exposure exercises allow the patient to develop new perspectives about the traumatic memory, learning that these reminders are safe. In addition to supporting treatment fidelity and patient engagement, the PE use case allows us to explore sensor-captured patient-generated data (sPGD) for patient homework review, mediating the space between self-report data and clinical intuition. Our work is guided by three  research questions:

\begin{itemize}
\item \textbf{RQ1:} What are veterans' and clinicians' perspectives on a clinician-facing interface to support intensive outpatient Prolonged Exposure therapy, a cognitive behavioral therapy for post-traumatic stress disorder? 
\item \textbf{RQ2:} What are potential challenges of integrating a clinical decision support system within the sensitive context of psychotherapy and veteran care?
\item \textbf{RQ3:} How can human-centered perspectives of Prolonged Exposure therapy provide further nuances into these complexities?
\end{itemize}

To address these questions and gain insights on our prototyped clinician dashboard, we conducted a two-phase interview study. We interviewed both PE clinicians (N=9) affiliated with, and former patients (N=7) of PE therapy who graduated from, a veterans' intensive outpatient program. Our key contributions are as follows. First, using thematic analysis, we distilled eight opportunities for a PE CDSS in the context of veteran care (e.g., offering homework review at a glance, aiding patient conceptualization, enabling ``detective work'') \textbf{(RQ1}; Sections \ref{findings_clinician}-\ref{findings_patient}), as well as three larger challenges related to context and deployment (e.g., understanding the role of the CDSS, navigating Veterans Affairs) \textbf{(RQ2}; Section \ref{findings_challenges}). Second, we reframe our findings to explore three human-centered perspectives on PE therapy as a sociotechnical system (distributed cognition, situated learning, and infrastructural inversion), offering design considerations and avenues of future work grounded in each \textbf{(RQ3}; Section \ref{discussion_alternative_views}). Our work echoes the importance of human-centered design. Rather than building a shallow understanding of a clinical context, human-centered approaches enable the distillation of richer and deeper insights by viewing the sociotechnical system from many different perspectives. 

\textbf{Content Warning:} This paper includes depictions of trauma and responses to traumatic experiences.

\section{Background}
\label{background}

\subsection{Veteran PTSD Care}
Post-traumatic stress disorder (PTSD) has an estimated lifetime prevalence of 6\% across all adults in the United States \cite{VaPtsdPrevalence}. PTSD is characterized by four symptom clusters: recurrent and intrusive experience of memories related to a traumatic event, persistent avoidance of trauma-related stimuli, negative alterations in mood and cognitions, and hyperarousal and reactivity \cite{american2013diagnostic}. Veterans are especially impacted by PTSD, as approximately 20\% of United States veterans of the conflicts in Iraq and Afghanistan meet the diagnostic criteria \cite{Cohen2010, Whealin2016}.  

Evidence-based psychotherapy (EBP) has been shown to be effective for treating PTSD  \cite{watkins2018treating}. In fact, the clinical practice guidelines released by the Department of Veterans Affairs (VA) and Department of Defense specifically recommend three trauma-focused EBPs for managing PTSD, including Prolonged Exposure therapy \cite{VA2023CPG}. Not many patients in the VA health care system, however, actually receive EBP: A national retrospective cohort study revealed that between 2001 and 2015, only 22.8\% veterans with PTSD who were receiving mental health care initiated EBP treatment and only 9.1\% completed treatment \cite{maguen2019factors}. The intensive outpatient format was devised to decrease this dropout rate through a condensed duration \cite{rauch2020prolonged,rauch2021intensive}. U.S. veteran mental health care is now largely subsidized by the Warrior Care Network \cite{VHAStats}, through which private organizations provide state-of-the-art intensive outpatient care to veterans for free. We dedicate the next section to one of these treatments: intensive outpatient Prolonged Exposure (PE) therapy.

\begin{table*}[]
  \caption{Description of PE Therapy for a Mock Trauma}
  \label{tab:PE_description}
  \renewcommand{\arraystretch}{1.2} 
  \centering
  \begin{tabular}{|p{45em}|}
    \hline
    \textbf{Clinical Session} (one-on-one):  \\
    Individual with PTSD recounts their trauma-related memory (e.g., \textit{an accident they witnessed at a playground}) and shares associated emotions and perspectives while also inhibiting all avoidant or self-soothing behaviors (e.g., bouncing leg, changing conversations). The clinician keeps track of a PE session checklist that includes 5 to 8 tasks for every session and 3 to 8 sub-tasks for every task. This includes having a patient rate their SUD score every 10 minutes.\\ \hline
    \textbf{Imaginal Homework Session}:  \\
    Patients are asked to sit still and listen to the recording of the clinical session; 
    Target: 30 minutes. \\ \hline
    \textbf{In Vivo Homework Session}: \\
    The patient completes at least one task in the In Vivo hierarchy related to \textit{the accident they witnessed at a playground}; Target: 30 minutes or until SUDs drop by 50\%.  \\ \hline
    \textbf{Example of an In Vivo Task Hierarchy} for a \textit{playground-related trauma} and associated SUDs \\
    SUD: \textbf{55} --- Drive around schools and parks early in the morning when there are no kids around \\ 
    SUD: \textbf{60} --- Park in front of a kids' playground early in the morning when there are no kids around \\ 
    SUD: \textbf{65} --- Walk around the perimeter of a park when there are no kids around \\ 
    SUD: \textbf{75} --- Sit on a park bench where a kids' playground can be seen at a distance when there are no kids around \\
    SUD: \textbf{85} --- Walk around a playground when there are children around \\  
    SUD: \textbf{95} --- Sit at a park bench among screaming and laughing children \\ \hline 
  \end{tabular}
\end{table*}

\subsection{Prolonged Exposure Therapy: A Cognitive Behavioral Therapy}
\label{background_PE}

Cognitive Behavioral Therapies (CBTs) are evidence-based psychotherapeutic approaches that aim to correct unhelpful avoidance and cognitive perspectives to address psychological or behavioral distress \cite{beck2009depression}. Between-session homework assignments are an integral part of CBT and enhance clinical outcomes \cite{mausbach2010relationship,kazantzis2010meta}. Some therapies under the CBT umbrella are also manualized, i.e., they have a strictly delineated session-by-session protocol. There is long-standing clinical evidence of the efficacy of manualized psychotherapies \cite{luborsky1993benefits, addis2000national, luborsky1984use, wilson1996manual, wilson1998manual,wilson2007manual}. Clinicians must implement the therapy according to the manual for high fidelity, although the manuals do advocate for some flexibility to prioritize patient needs, for example in cases of suicidal or homicidal behavior \cite{foa_prolonged_2019}. Despite the strong research support, the access and utilization of manualized CBTs are hindered by extensive training requirements, therapist burden, and implementation difficulties \cite{rosen2016review, shiner2013measuring}. 

Prolonged Exposure (PE) therapy is an evidence-based manualized CBT  \cite{bisson2013psychological, rothbaum2002exposure} for treating PTSD. PE utilizes experiential exercises during and between sessions to teach patients how to gradually approach their traumatic memory and reminders of this memory without avoidance. These exposure exercises allow the patient to develop new perspectives about the traumatic memory, learning that these reminders are safe and do not need to be avoided. PE clinicians follow a strict session-by-session protocol, detailing one-on-one clinical sessions with a psychotherapist as well as between-session homework activities. In an intensive outpatient program (IOP), PE is delivered via daily sessions for two weeks, a formulation that heavily reduces dropout rate \cite{rauch2020prolonged,rauch2021intensive}. \textbf{Imaginal exposure} is first conducted during a one-on-one clinical session, during which an audio recording is collected of the patient describing the narrative of their traumatic memory. Subsequent imaginal exposures are then completed as homework by listening to this audio recording while preventing all distractions. \textbf{In vivo exposure} involves gradually approaching real-world stimuli that remind the patient of their traumatic event and therefore cause distress. In vivo exposures are typically executed as homework exercises, and patients self-report their level of distress through Subjective Units of Distress (SUDs). SUDs range from 0 to 100, with 0 representing no negative affect and 100 representing unbearable stress. See Table \ref{tab:PE_description} for an overview of the components of PE therapy and examples of in vivo exposures.

\section{Related Work}

\subsection{Technological Support of Cognitive Behavioral Therapies}
\label{related_work_TechForCBT}

The access and utilization of cognitive behavioral therapies (CBTs) remain low despite the large demand for mental health treatments \cite{kazdinAddressingTreatmentGap2017a, blane2013cognitive, coyleComputersTalkbasedMental2007, oconnorIncreasingAvailabilityPsychological2018}. There is currently a shortage of trained psychotherapists \cite{oconnorIncreasingAvailabilityPsychological2018, blane2013cognitive} due in part to costly and labor-intensive training methods \cite{oconnorIncreasingAvailabilityPsychological2018, fairburn2011therapist, martino2016effectiveness}. Researchers in human--computer interaction (HCI), computer-supported cooperative work (CSCW), and health informatics have turned to technology to support CBT for mental health conditions \cite{aguilera_theres_2014, sanches_hci_2019} including depression \cite{rohani_mubs_2020, rohani_personalizing_2019, kawanishi_lifelog-based_2015, schueller_initial_2015}, insomnia \cite{zhu_making_2017}, and PTSD \cite{kuhn_randomized_2017, reger2013pe}. 

Telehealth and mobile health offer opportunities to address the low access and utilization of CBT. Largely self-guided CBT delivered via the internet has been explored, but had lower clinical efficacy and higher attrition than in-person treatment \cite{kuester2016internet, sijbrandij2016effectiveness}. Remotely delivered CBT via video conferencing software has been shown to be as effective as in-person CBT for high-intensity intervention for a variety of mental health conditions \cite{schueller_initial_2015, maieritsch2016randomized, bower2005stepped}. Researchers in mobile health have proposed self-contained mobile systems informed by CBT  \cite{sanches_hci_2019} to support conditions including social anxiety \cite{rennert_faceit_2013} and stress \cite{chow_feeling_2024}. Rohani et al., for example, developed mobile systems to support the behavioral activation component of CBT by assisting patients plan, practice, and reflect  \cite{rohani_personalizing_2019} and recommending activities based on mobile sensors \cite{rohani_mubs_2020}. 

The ubiquity of mobile and wearable devices has enabled researchers to further support CBT by augmenting self-report measures from clinical homework (an essential part of many CBTs \cite{wenzel2017basic}). Self-report data is inherently biased \cite{foster_validity_1995} and could miss important details. Researchers have therefore explored self-tracking and sensors to support patients in collecting data during homework activities \cite{kawanishi_lifelog-based_2015, matthews_mood_2011, reger2013pe}.  Previous research suggests the potential of sensor-captured patient-generated data (sPGD) collected in the wild to support CBT homework review \cite{ng2019provider, evansUsingSensorCapturedPatientGenerated2024} and facilitate data-driven communication between clinicians and patients \cite{mishra2018supporting, schroeder_supporting_2017}. sPGD may also empower patients to take a more active role in their health management \cite{evansUsingSensorCapturedPatientGenerated2024,khatiwada2024patient}, especially when the data streams validate their lived experiences \cite{ng2019provider}. Taking advantage of various sensors on smartphones or wearables, sPGD can provide potentially useful contextual information on patients outside the clinic \cite{neff2016self, west_common_2018, bardram_decade_2020}, enabling a movement towards ``technology-supported continuous mental healthcare'' \cite{bardram_decade_2020}. Previous work has suggested that sPGD can be used to infer the mental health status of patients with conditions such as depression \cite{braund_smartphone_2022, kelley_self-tracking_2017, wang_tracking_2018, rohani_mubs_2020}, bipolar disorder \cite{braund_smartphone_2022, frost_supporting_2013, bardram_designing_2013, bardram_designing_2016, rohani_mubs_2020}, and eating disorders \cite{juarascio_momentary_2020, vega_detecting_2022}. 

Rather than following previous explorations that aimed to support patient engagement and adherence, this work instead focuses on clinician-facing technology (i.e., a CDSS) to support the delivery of a manualized CBT. Further, we engage both practicing PE clinicians and former PE patients (veterans of the United States military) to better understand their experiences of PE therapy and perspectives on a CDSS for this context. Based on interviews with both participant groups, we provide an overview of perceived roles that a CDSS could play in therapy practice (Sections \ref{findings_clinician}-\ref{findings_patient}), including offering homework review at a glance, aiding patient conceptualization, and enabling ``detective'' work.

\subsection{Designing Clinical Decision Support Systems (CDSSs) for Sensitive Contexts }
\label{related_work_DesigningCDSSs}

Clinical decision support systems (CDSSs) have attracted the attention of researchers as a promising way to facilitate the clinical workflow. CDSSs may provide informational and computational assistance for diagnosis \cite{beede_human-centered_2020, wang_brilliant_2021}, medication prescription \cite{jo_designing_2022, jacobs_designing_2021}, patient evaluation \cite{lee_towards_2022, lee_human-ai_2021}, problem-based discovery \cite{zhang2022get}, and sense-making \cite{jin2017collaborative}.

Successful design and deployment of a CDSS requires a deep understanding of the context in which it will be deployed. CDSSs have not yet been successfully adopted in real-world practice \cite{middleton_clinical_2016, yang_unremarkable_2019, kouri_providers_2022, zhang2022get}, largely due to a lack of sufficient consideration of user needs and contextual fit \cite{ yang_investigating_2016,  cai_hello_2019, khairat_reasons_2018, wang_human-centered_2023, zhang2022get, jo_designing_2022}. For example, concerns of sensor-captured patient-generated data (sPGD) include unclear benefits to patients \cite{bardram_decade_2020, ng2019provider}, the time-consuming use of data at the point of care \cite{ng2019provider, evansUsingSensorCapturedPatientGenerated2024}, and conflicts with existing clinical workflows \cite{arean2021using}. AI-driven CDSSs similarly raise sociotechnical challenges including technical limitations,   informational barriers, workflow misalignment, attitudinal barriers, usability issues, and environmental barriers \cite{wang_human-centered_2023}. 

The importance (and complexities) of contextual fit are especially salient when considering sensitive clinical contexts. Designers have taken a variety of approaches when unable to directly observe real-world clinical interactions. Many have involved end-users in exploratory, qualitative studies (e.g., interviews) to better understand their needs,  challenges, and workflow \cite{leist2025towards,bhattacharya2023directive,thieme_designing_2023,cai_hello_2019,jo_designing_2022}. Others involved stakeholders throughout the design process, for example through co-design or partnerships with clinical experts  \cite{jacobs_designing_2021, zhu2025designing,javaheri2025concept}. Approaches such as contextual design and ethnography have offered broader and more complex pictures of clinical contexts \cite{kluber2020experience,beede_human-centered_2020}, including environmental factors of which primary stakeholders may be unaware.  Still, these approaches may not necessarily facilitate alignment with accepted clinical protocols. While there do exist CDSSs designed in accordance with specific clinical standards (e.g., a CDSS that operationalizes a clinical practice guideline for opioid therapy \cite{trafton2010designing}), few HCI studies in recent years have directly leveraged clinical protocols when attempting to understand a design context. Exceptions include a conversational AI therapist based on frameworks of CBT and motivational interviewing \cite{nie2024llm}, and a voice assistant that follows the session flow and activities outlined in the cognitive simulation therapy manual \cite{qiu2025voice}.

While recent work has explored CDSSs for mental health \cite{golden2024applying,tong2025clinical}, including sPGD to support clinical decision making (see Section \ref{related_work_TechForCBT}), the specific context of veteran care introduces additional considerations for CDSS design. Prior work highlights the value of leveraging inputs from trusted others in veteran PTSD care \cite{evans_perspectives_2022,schertz_bridging_2019}, veteran preferences for sharing their health data \cite{hogan2025veteran}, and the importance of considering how military identity can influence behaviors and attitudes towards care \cite{evans_understanding_2020}. For example, military culture may foster priorities of hyper-masculinity, stoicism, and focusing on others \cite{rauch_expanding_2017,semaan2017military}, as well as an ability to ``turn off'' emotions to keep calm \cite{zwiebach2019military}. These considerations also drive veteran-centered design considerations for other technologies, including peer support interventions \cite{addis2000national} and support seeking in online communities \cite{zhou2022veteran}.

Concerns of technology integration into clinical workflows indicate persisting knowledge gaps in understanding how to design CDSSs to facilitate successful deployment and integration. The sensitive nature of psychotherapy (especially for veterans) precludes designers from direct observation of this design context. Manualized CBT, however, offers an opportunity to directly leverage clinical manuals in the CDSS design process. We designed a Figma prototype of a clinician dashboard to support Prolonged Exposure (PE) therapy treatment for veterans with PTSD, relying heavily on the PE protocol \cite{foa_prolonged_2019}. Our findings include three challenges regarding contextual fit that arose during interviews with PE therapists (assisted by this technology probe) and graduates of a veterans' PTSD treatment program (Section \ref{findings_challenges}). These challenges are related to the role the CDSS should play in patient-clinician interactions, differing perspectives on psychophysiology, and the complexities that arise due to the Department of Veterans Affairs (the VA).

\subsection{Considering Sociotechnical Lenses for System Analysis}
\label{related_work_ViewingSystems}

Human-centered computing researchers are likely familiar with the benefits of leveraging theoretical and analytic lenses to better understand a sociotechnical system. The field of HCI itself has rapidly expanded by incorporating approaches and theories from a wide variety of disciplines \cite{rogers2012hci}. Taking this broader sociotechnical perspective is especially important in health contexts, even for a CDSS that anticipates only a single user \cite{marathe2021situated}. Below we introduce three perspectives (distributed cognition, situated learning, and infrastructural inversion) that have previously been applied to clinical domains in HCI. We return to these three perspectives in Section \ref{discussion_alternative_views}.

Hutchins proposed distributed cognition theory to analyze how both internal and external representations of information (via actors, artifacts, etc.) contribute to an overall information flow within a sociotechnical system \cite{Hutchins1995}. Distributed cognition has been widely applied to clinical settings, providing insights into team coordination \cite{wilson2023scoping} and impacts of electronic medical record design \cite{wilson2023scoping}. Numerous attempts have been made to expand upon distributed cognition theory to transition from analysis to actionable design implications. Distributed Cognition for Teamwork (DiCoT) \cite{Blandford2005,Furniss2010} has made the largest inroads into complex health systems, identifying barriers to information flow and collaboration in contexts such as Emergency Medical Services \cite{Zhang2021}, the intensive care unit \cite{hussain2016can}, and emergency cardiac arrests \cite{Morand2022}. In Section \ref{discussion_dicog}, we discuss how a distributed cognition perspective can be beneficial in understanding the flow of information during PE therapy, a sociotechnical system involving many internal and external components.

Lave and Wenger \cite{lave1991situated} provided an analytic perspective on situated learning. Situated learning views learning as social and inseparable from context, rather than as simple knowledge transfer from teacher to student. Situated learning and the process of legitimate peripheral participation \cite{lave1991situated} have been utilized in HCI and CSCW for online health communities and safe spaces \cite{nova2024unveiling,tam2023learning} and the online experiences of community health workers \cite{ismail2019empowerment}.  Psychotherapist trainees can be seen as learners, both in formal clinical education \cite{feinstein2021descriptions} and less formal settings (e.g., volunteer counselors \cite{yao2022learning}). Psychotherapy may also be treated as a form of learning in itself, positioning patients, too, as learners \cite{bandura1961psychotherapy,rose2005counselling}. Situated learning may be a particularly insightful lens for PE therapy, given the centrality of in vivo homework exercises that emphasize patient learning in real-world contexts. In Section \ref{discussion_learning}, we highlight the utility of a situated learning perspective on CBT in probing questions of power dynamics and access to learning.

Infrastructure describes the systems and mechanisms that are ``by definition invisible, part of the background for other kinds of work'' \cite{star1999ethnography}.  Infrastructural inversion is an analytical approach that intentionally pulls this background to the foreground \cite{bowker1994science}: ``a struggle against the tendency of infrastructure to disappear'' \cite{bowker2000sorting}. An inherently critical process \cite{hetherington2018infrastructure}, HCI and CSCW researchers have leveraged infrastructural inversion to reveal insights in domains such as climate and disaster risk modeling \cite{paudel2025hype}, health management \cite{xu2024obviously}, and hospital coordination \cite{simonsen2020infrastructuring}. While this is frequently accomplished by exaggerating and interrogating points of breakdown, infrastructural inversion can also offer generative opportunities \cite{wood2017lies,simonsen2020infrastructuring} for more disruptive technologies and critical practices \cite{paudel2025hype}. Infrastructural considerations are especially vital in clinical settings, in which infrastructure is ``a key pillar supporting the fundamental aim of promoting improved standards of care'' \cite{luxon2015infrastructure}. Recent work in health HCI has highlighted the complex (and yet frequently invisible) ``infrastructure work'' that goes into building and maintaining infrastructure for mental health peer support platforms  \cite{ding2023infrastructural}, and how individuals navigate infrastructure for health self-management \cite{xu2024obviously}.  In Section \ref{discussion_infrastructure}, we reframe the PE manual as a key aspect of PE infrastructure, and discuss the ways in which introducing a CDSS as a technological intermediary might drive infrastructural inversion. 

These brief overviews of distributed cognition, situated learning, and infrastructural inversion certainly do not cover the full breadth of work that leverages these lenses within HCI and related fields; however, they hint at important intersections within the context of intensive outpatient psychotherapy for veterans with PTSD. This is an information-heavy clinical setting yet also a learning context, heavily influenced by the organizational aspects of veteran care. In Section \ref{discussion_alternative_views}, we reframe our findings to discuss how these perspectives might differently drive the design of a CDSS for veteran PTSD care. 
\section{Methodology}

We conducted a two-phase interview study to gain insights into a novel clinical decision support system (CDSS) to support the practice of Prolonged Exposure (PE) therapy for veterans. Both PE clinicians (N=9) and former patients of PE therapy (N=7) were interviewed, and the former interacted with a Figma prototype of a clinician dashboard. Below we detail our positionality, the prototype design process, the interview study, and our data analysis approach.

\subsection{Ethical Disclosure \& Positionality}

The collection of patient data and the incorporation of technology into clinical practice raise potential privacy, security, and ethical concerns. This includes skepticism regarding physiological sensor data, automation bias due to quantitative data, and concerns of dehumanizing spaces that are typically reserved only for human-human interaction. This work is exploratory, intending to provide insights into opportunities for a CDSS and the implications for integration into sensitive psychotherapy settings.

We disclose our positionality to situate how our study was conducted and data interpreted. Our team includes researchers from a variety of disciplines (human-centered computing, computer science, engineering, philosophy, and psychology). It also includes a clinical psychologist who has an active treatment practice for individuals with PTSD and leads training sessions for clinicians who aim to become PE practitioners. Our authorship includes representation from multiple mental health conditions, which informed our decision to include perspectives not only from PE clinicians but also from individuals who have completed PE therapy. Our authors' approaches to technology vary, ranging from techno-pragmatists to techno-skeptics. We take a human-centered approach to better understand the complex dynamics and challenges within this sensitive sociotechnical system, rather than simply advocating for the integration of technology.

\subsection{Prototype Design}

\subsubsection{Design Approach}

We were precluded from first-hand observation of patient-clinician interactions due to the sensitive nature of the clinical context \cite{yang_unremarkable_2019}. Instead, the initial dashboard was designed based on previous work \cite{schertz_bridging_2019, evansUsingSensorCapturedPatientGenerated2024}, the clinician guidebook for PE \cite{foa_prolonged_2019}, and the clinical perspective of our co-author who has extensive experience practicing PE therapy, supervising novice PE clinicians, and conducting PE therapy research. 

For approximately one year, the Figma dashboard prototype was iteratively designed and improved by the interdisciplinary research team during weekly sessions. De-identified patient cases were shared during these meetings, and patient symptoms, clinical homework, and interpersonal interactions were discussed to understand how the CDSS might support (or hinder) PE delivery in specific cases. To ensure that the design decisions were sound, two other PE experts were consulted during monthly project meetings.

\begin{figure*}[h!]
    \centering
    \includegraphics[width=0.88\linewidth]{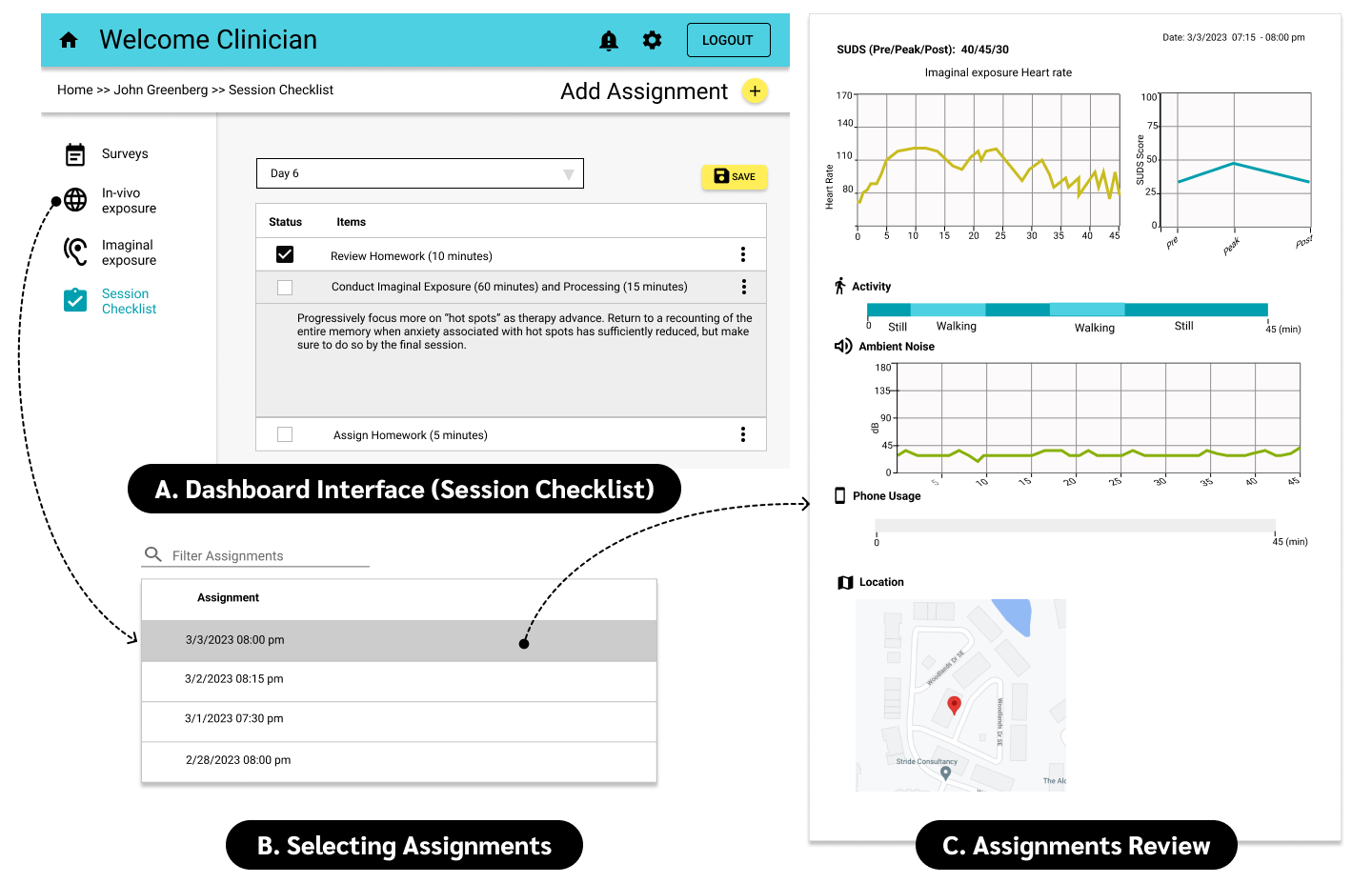}
    \caption{Screenshots from the Figma prototype of a clinician dashboard to support Prolonged Exposure therapy. The Session Checklist page provides checklist items from the clinician guidebook \cite{rauch2020prolonged, foa_prolonged_2019} and allows clinicians to edit checklist items as needed \textbf{(A)}. The in vivo exposure pages provide a list of assignments finished by the patient \textbf{(B)} and details about a chosen homework submission including SUDs and sPGD \textbf{(C)}.}
    \Description{Screenshots from the Figma prototype of a clinician dashboard to support Prolonged Exposure therapy. The first screenshot is of the Session Checklist page. This provides checklist items from the clinician guidebook (e.g., "Review Homework (10 minutes)", "Conduct Imaginal Exposure (60 minutes) and Processing (15 minutes)") and allows clinicians to edit checklist items as needed. The second screenshot is of a drop down search bar which allows clinicians to select an in vivo assignment finished by the patient, sorted by date and time. The third screenshot shows details of a chosen homework submission, including a graph of exposure heart rate over time, SUDs score over time, activity throughout the exposure ("still" or "walking"), ambient noise level over time, phone usage over time, and a GPS pin on a map}
    \label{fig:screenshots}
\end{figure*}

\subsubsection{Prototype Overview}
\label{datastreams_here}
The prototyped dashboard (see Figure \ref{fig:screenshots}) is intended to support adherence to the clinical manual and assist in homework review, including sensor-captured patient-generated data (sPGD). The three overall goals of the CDSS are to support fidelity to the clinical protocol, assist clinicians in the mediation between patient self-report and clinical intuition, and enhance patient engagement. We aimed to design a CDSS that would support the existing manualized protocol, rather than replace this workflow and interfere with therapy fidelity. Prior literature highlights the importance of considering the impact on psychotherapists' workflow when introducing new technology \cite{simionatoBurnoutEthicalIssue2019, barnettEthicalPracticePsychotherapy2019}. The utility of a CDSS is known to reduce when technological interventions hamper clinicians' usual workflow \cite{jo_designing_2022, jacobs_designing_2021, yang_investigating_2016}.

\begin{table*}[h] 
\centering
\caption{Participant Information for Nine PE Clinicians}
\label{tab:participants} 
\renewcommand{\arraystretch}{1.2}
\setlength{\tabcolsep}{4pt}
\begin{tabular}{|c|cccccc|}
\hline
 &  &  & Clinical & PE & Telehealth & Supervision \\
ID & Gender & Age & Experience & Experience & Experience &  Experience \\
 & & & (years) & (years) &  (Y/N) &  (Y/N) \\ \hline
\textbf{C1} & Male & 50 - 59 &15 - 20  & 15 - 20 & Y  & Y\\ \hline
\textbf{C2} & Male & 50 - 59 &20 - 25  & 10 - 15 & N & Y   \\ \hline
\textbf{C3} & Female & 40 - 49 &20 - 25 & 20 - 25 & Y & Y  \\ \hline
\textbf{C4} & Female & 60 - 69 &35 - 40 & 35 - 40 & Y & Y  \\ \hline
\textbf{C5} & Male & 30 - 39 & 5 - 10 & 5 - 10 & Y & N   \\ \hline
\textbf{C6} & Female & 30 - 39 & 5 - 10 & < 5 & Y & Y  \\ \hline
\textbf{C7} & Male & 30 - 39 & 10 - 15 & 10 - 15 & Y & Y  \\ \hline
\textbf{C8} & Female & 30 - 39 & 5 - 10 & < 5  & Y & N  \\ \hline
\textbf{C9} & Female & 30 - 39 & 5 - 10 & < 5  & Y & N  \\ \hline
\end{tabular}
\end{table*}

The clinician dashboard prototype includes four key components, which follow the PE protocol \cite{foa_prolonged_2019}:  \textbf{surveys} overview, \textbf{in vivo exposure} page, \textbf{imaginal exposure} page, and \textbf{session checklist}. The \textbf{surveys} page shows an overview of results from surveys the patient has taken, e.g., PCL-5 \cite{blevins2015posttraumatic} for PTSD-related symptoms. The \textbf{in vivo exposure} and \textbf{imaginal exposure} pages display the sPGD collected during homework, as evidence of patient engagement or avoidance behaviors (described in the next paragraph). The digitalized \textbf{session checklist} was included because the PE manual \cite{foa_prolonged_2019} states that the checklists should be used in all sessions to ensure high fidelity. Aligned with Shneiderman's ``Overview first, zoom and filter, and details on demand'' \cite{shneiderman2003eyes}, all pages provide an overview chart  to show the overall progress during the therapy, allow clinicians to filter results based on a keyword, and reveal details about clinical homework or a survey submission when clicked. See Figure \ref{fig:screenshots} for prototype screenshots. 

The sPGD included for imaginal and in vivo exposures were chosen based on our clinical co-author's expertise and previous mental health studies utilizing sensor data \cite{bardram_designing_2013, frost_supporting_2013, wang_crosscheck_2016, wang2014studentlife, schertz_bridging_2019, evansUsingSensorCapturedPatientGenerated2024}. To make the prototype more realistic, mock sPGD was generated with our clinical co-author's guidance. Below, we describe how different sPGD data streams might be used during PE therapy:
\begin{itemize}
\item \textit{Heart Rate:} Heart rate is expected to increase when emotionally processing and decrease with habituation. If heart rate associated with a distressing stimulus decreases over the homework sessions, this suggests progress. A stable heart rate during emotional processing may indicate avoidance.  
\item \textit{Phone Usage:} Patients are asked to focus on their homework exposure. Some patients may find this very stressful and emotionally disengage from homework by using smartphone applications, making phone calls, or texting.  
\item \textit{Physical Activity:} During imaginal exposures, patients are expected to sit still and focus on an audio recording of their trauma narrative.  Some in vivo exposures, in contrast, intentionally involve physical activity, e.g., walking down a street. Pacing or sitting still, respectively, may indicate avoidance.  
\item \textit{GPS:} Patients are expected to complete imaginal exposures at home, while in vivo exposure may involve visiting a specific place. GPS could thus provide contextual information and confirmation about homework setting.
\item \textit{Ambient Noise Level:} Similar to GPS, ambient noise level could provide contextual information regarding homework requirements. For example (from Table \ref{tab:PE_description}), one would expect a high level of noise in a playground if it is full of children.
\end{itemize}

\subsection{Study Recruitment \& Participation}

Our university IRB approved the interview protocol, and all participants provided consent by digitally signing consent forms before beginning their interview.

We recruited nine \textbf{PE clinicians} (denoted C1-C9) via word of mouth or a recruiting email from our clinical co-author. We required clinicians to have prior experience practicing PE therapy, but purposefully recruited clinicians with varying levels of PE experience (see Table \ref{tab:participants}). All clinicians have a doctoral degree in clinical psychology and are affiliated with a private institution's veteran-specific PE intensive outpatient program. Some of the clinicians hold a dual appointment at the Department of Veterans Affairs (VA). Because of the impact of the COVID-19 pandemic and efforts to make PE therapy more accessible, six of the clinicians provide PE in telehealth settings. Six have supervised other clinicians and two had a supervisor at the time of the interview. Three of the most experienced clinicians are part of nationally recognized PE training programs that have trained over 100 mental health workers. We compensated each clinician with \$20 cash or a gift card for one 60- to 90-minute interview. 

In addition, we interviewed seven \textbf{former PE patients} (denoted V1-V7) who were all veterans of the U.S. military. These former patients were graduates of an intensive outpatient veterans' treatment program (the same program with which the clinicians are affiliated) and were recruited by our clinical co-author. We do not report their demographics nor the time of their treatment to protect their anonymity due to the sensitive nature of their treatment.

\subsection{Data Collection \& Analysis}

Individual semi-structured interviews with the nine \textbf{PE clinicians} were conducted via Zoom. Participants were first made aware that we aimed to understand their current PE practices and how technology might support their work. We emphasized that the CDSS was not intended to change or replace their existing practice, exercising caution because a technology probe approach can suggest a techno-positivist (non-neutral) stance. Clinicians shared their current workflow, any technology use during their therapy practice, how they determine patient progress, and the challenges they face in delivering PE therapy. They then engaged in a think-aloud \cite{jaaskelainen2010think} session with the clinician dashboard prototype. Clinicians were requested to interpret the mock sPGD as if it belonged to one of their patients. They provided feedback on the dashboard design, including usefulness of the visualizations and potential impacts on their practice.  

Semi-structured interviews with the seven \textbf{former PE patients} were conducted via Zoom. The patients were asked about their experiences completing PE homework activities (e.g., exposure exercises), evaluating their engagement and progress, and reporting their experiences during appointments. They were also asked to imagine and discuss a patient-facing smartphone application that would collect sPGD (the same data streams as described in Section \ref{datastreams_here}) to provide insights into their homework experiences and level of engagement. Former patients were told that ``data streams including heart rate, physical activity, location, phone usage, and noise level in the environment'' would be displayed ``both [on] a patient-facing mobile application and a clinician dashboard.'' 

All interviews were recorded, automatically transcribed, and manually reviewed for accuracy. In a two-phase approach, we qualitatively analyzed the transcripts using bottom-up, reflexive thematic analysis \cite{braun2006using,braun2019reflecting}, aiming to identify ``patterns of shared meaning'' \cite{braun2019reflecting}. First, the second co-first author open-coded only the clinician interview transcripts to identify patterns and themes, and the authors discussed the themes. In the second phase, the first co-first author analyzed the patient interviews, re-analyzed the clinician interviews, and authors came to consensus on updated themes. Our clinical co-author also reviewed the final themes to validate that our interpretation is aligned with their clinical expertise.

\section{Findings}

\begin{table*}[] 
\centering
\caption{Reflexive Sub-themes}
\label{tab:themes} 
\renewcommand{\arraystretch}{1.2}
\setlength{\tabcolsep}{4pt}
\begin{tabular}{|p{16em}|p{28em}|}
\hline
\textbf{Section} & \textbf{Sub-theme} \\ \hline

\multirow{3}{16em}{Aiding Protocol Fidelity and Clinician Workflows (Section \ref{findings_clinician}) }&  By-the-Book Informational Support \\ \cline{2-2}
&  Streamlined Recording Keeping and Homework Data at a Glance\\ \cline{2-2}
& The Longer Journey: Ease of Data Transfers \\ \hline

\multirow{2}{16em}{Mediating Patient Self-Report and Clinical Intuition (Section \ref{findings_mediation}) }&  Conceptualization of Patients: Context and Values \\ \cline{2-2}
& ``Detective'' Work with Self-reports and Data Streams \\ \hline

\multirow{3}{16em}{Supporting and Engaging Patients (Section \ref{findings_patient})} & Reducing the Distance from Me to My Clinician \\ \cline{2-2}
& Achievements for the Fridge \\ \cline{2-2}
&  Unpacking the Nuances of the ``PTSD Veteran'' Identity \\ \hline

\multirow{3}{16em}{Contextual Challenges for CDSS Design (Section \ref{findings_challenges}) }&  Understanding the Role of the CDSS: The ``In Between'' \\ \cline{2-2}
& Reaching a Community Consensus on Psychophysiology \\ \cline{2-2}
& The Elephant in the Room: Navigating Veterans Affairs \\ \hline

\end{tabular}
\end{table*}

From our interviews with nine Prolonged Exposure (PE) therapy clinicians (C1-C9) and seven former PE patients (V1-V7), we distilled perceived opportunities for a clinical decision support system (CDSS) to facilitate the delivery of psychotherapy for veteran PTSD care. We discuss these opportunities in three sections: supporting clinician workflow and fidelity to the protocol (Section \ref{findings_clinician}), mediating patient self-report and clinical intuition (Section \ref{findings_mediation}), and supporting and engaging patients (Section \ref{findings_patient}). In Section \ref{findings_challenges}, we then detail three challenges related to deploying a CDSS in this context. Our eleven reflexive sub-themes are listed in Table \ref{tab:themes}.

Per \textbf{RQ1}, we aimed to gain veteran and clinician perspectives on a CDSS for this use case. As such, only some of the findings below directly relate to the veteran context: Many of the veterans expressed opinions that they did not explicitly contextualize within their military identity.

\subsection{Aiding Protocol Fidelity and Clinician Workflows}
\label{findings_clinician}

\subsubsection{By-the-Book Informational Support}
\label{opportunity_info_support}

Because of the demands of a manualized CBT, informational support is helpful in ensuring fidelity to the protocol. Most of the clinicians we interviewed use materials from the PE manual such as the session checklist \textit{``to make sure that I've gotten through everything important''} (C3). C1 said that \textit{``I know the session pretty well at this point, but I still bring the outline to the session. It helps us deliver the treatment consistent with how it's designed.''} Newer clinicians in particular rely on the checklist page and even include customized versions with additional notes or rationale. C8, for example, frequently pulls \textit{``some of the stuff from the manual of like, [...] you're conducting imaginal exposure [...], I can like briefly look at that [...] and be like, `Okay, this is the metaphor we use.'\thinspace''} C2 also emphasized the importance of informational support for clinicians affiliated with Veterans Affairs because \textit{``you might only have a couple of patients at a time who are doing PE because of the constraints on the system [....] So it might be that [...] you would need a little refresher.''} 

A CDSS that mimics the materials already utilized by clinicians was well-received, especially for telehealth settings; however, the importance of informational support varied depending on a clinician's specific approach. Clinicians commonly aimed to \textit{``deliver the treatment consistent with how it's designed''} (C1), but some referred to the protocol more as a guide rather than gospel. C5, for example, when explaining specifics of their approach, said \textit{``I think that's what the manual says... And if not, that's kind of how I've internalized it.''} The perceived benefit of a CDSS providing informational support therefore varied: Clinicians new to PE and those committed to rigid ``by the book'' therapy delivery were optimistic, while others would view this informational support more as \textit{support} rather than strict direction.

\subsubsection{Streamlined Recording Keeping and Homework Data at a Glance}
\label{opportunity_homework_review}

Clinicians felt that the dashboard prototype would offer better access to patient data collected during homework exposures. Some patients took \textit{``notes like on a piece of paper and [...] just talked off of the notes''} during the clinical session, but this consumes valuable time. Cumbersome homework review was a weakness C9 pointed out for PE coach \cite{reger2013pe}, the popular companion mobile application for PE patients. C9 noted that there is \textit{``no clinician version [of PE Coach]. [...] If we were in person, they would just show me on their phone, but in case of a telehealth setting [...] there’s no way you can access their data.''} A clinician interface with patient data ``at a glance'' could facilitate data sharing by allowing clinicians to quickly review a patient's homework notes before their appointment. This quick patient data review could {also} help facilitate discussions between newer clinicians and their supervisors. From the supervisor perspective, C4 suggested that a CDSS could aid therapist-supervisor discussions around patient learning, as one of their \textit{``favorite questions to ask younger therapists is, `What does the patient need to learn?'\thinspace''} From a trainee perspective, as C8 noted, \textit{``I could say, `hey, I'm concerned because their in vivo [...] isn't going down and they've done it five times. Look at this pattern and help me see what I'm missing.'\thinspace''} 

Opportunities to facilitate data review extended to clinician-recorded data. Because their electronic medical record (EMR) systems do not support visualization, some clinicians use external spreadsheets to monitor patient progress, make informed decisions, and facilitate conversation with the patient. C5 noted, however, that \textit{``to have it all in one place, might be my biggest gripe.''} A CDSS might ease this burden by gathering all of these records in one place and automatically generating visualizations with clinically relevant characteristics. C3 believed this could aid clinical decision-making, \textit{``to help figure out what that next step would be for the patient.''} 

\subsubsection{The Longer Journey: Ease of Data Transfers}
\label{opportunity_transfers}

As is expected for systems that target integration within a work environment, a CDSS would need to ease clinical workflows, or at least not negatively impact them. Patients and clinicians both believed that the prototyped clinician dashboard could facilitate clinical information transfer. This would be helpful when a clinician is asked to cover for a colleague (for example due to illness or vacation), essentially \textit{``jumping in and doing like one day of therapy with a person that I had never met. [...] And so it might be nice [...] to kind of show where they're at in treatment''} (C8). Patient concern regarding the subjectivity of data interpretation was a caveat, however, as a new provider's \textit{``opinion of how your session went may differ from the other provider''} (V1). 

Patients also envisioned their data continuing with them throughout their larger mental health journey with \textit{``a different organization or different psychologist,''} as patients \textit{``keep building on my [...] tolerance, my knowledge [...] more tools for my toolbox''} (V3). Multiple patients expressed that they would want access to all of their sPGD data, \textit{``to be able to review it, share it [...] with another provider, and showcase that to them''} (V7).

\subsection{Mediating Patient Self-Report and Clinical Intuition}
\label{findings_mediation}

\subsubsection{Conceptualization of Patients: Context and Values}
\label{opportunity_conceptualization}

Both clinicians and patients believed that a CDSS would aid clinicians in building improved mental models of patients. C3, for example, felt that it would provide insights into how to engage a patient, especially \textit{``for patients who are having trouble, this could be another really useful piece of information to help [...] pull for more emotion.''} Participants mentioned two additional areas in which they imagined the prototyped clinician dashboard could aid patient conceptualization.

First, participants felt that the CDSS could provide additional context on how patients report their experiences. C7 mentioned a case where a patient continued to report high distress throughout therapy, which turned out to be because \textit{``he was like really unhappy in his job [...] and he wasn’t doing a good job of teasing that apart.''} C5 further stressed the importance of synthesizing heterogeneous data to figure out \textit{``what's driving them to report the way that they're reporting. [...] Are there other things [...] that I need to think about here?''} In fact, V1 was in support of data transparency, to prove to their clinician that they were telling the truth: \textit{``I'd rather them know who I am, by being able to use the data to determine whether or not I was using some fake data to say I had problems.''}

Participants noted that contextual data and sPGD collected during homework exposures could also help clinicians better understand patient priorities and values. Although the expectation is to ignore phone calls or texts during exposures, V6 said that \textit{``I'm just going to answer a phone call if it's important. And if it's somebody I know, like my wife, my son [...], but that's it.''} While a clinician might at first be surprised or dismayed to see that their patient answered a phone call during an exposure, this may then facilitate conversation to understand patient values more holistically. Similarly, the CDSS data may make cultural differences between clinician and patient more salient. C7 shared that they had a patient who was completing in vivo exercises but was not reporting lower distress. Although at first \textit{``it was very confusing, [...] this was a patient who was a Black man and he was going to a store to wander around [...] but he was being followed around by security. That is objectively distressing [...]. Like, okay, so we need to figure out another store or we need to [...] let them know that you're a veteran and you're doing an exposure.''}

\subsubsection{``Detective'' Work with Self-reports and Data Streams}
\label{opportunity_detective} 

Subjective, emotional experiences and physiological data \textit{``can be discordant [...] in the reporting''} (C2), leading to a patient and their clinician interpreting therapy progress differently. If a patient reports unchanged distress even though they seem to be making progress, this may lead to discussion to determine if they are \textit{``a poor reporter and we need to help them notice when things are changing, or if what I think I'm seeing is not what their internal experience is''} (C3). Discrepancies between internal (i.e., patient self-reported experience) and external  (i.e., sPGD on physiological or contextual factors) are therefore opportunities to facilitate clinical inquiries and deeper discussion. Still, most clinicians held the internal and subjective as paramount, respecting a patient's lived experience. C5 noted that \textit{``curiosity''} and collaboration is vital in exploring discrepancies \textit{``because this is their affective response. This is their emotional sphere.''}

When the therapy isn't as effective as expected, clinicians believed they could play a \textit{``little bit of a detective''} (C9) using these discrepancies. C9 gave the example that if a patient is \textit{``not habituating when they see a movie with a stabbing,''} the data might uncover that \textit{``it's because they were on Spotify the whole time.''} V1 noted, however, that this detective work might uncover things a patient doesn't necessarily want uncovered, for example if before an exposure, \textit{``you ate a medicinal brownie, [...] your results are going to be completely different. And then you get asked `How did you change your results?' You're not going to say `Oh, I ate a medicinal brownie.'\thinspace''} This suggests that although the CDSS could facilitate clinician-patient communication and teamwork, limitations exist. This benefit depends on existing trust within the therapeutic alliance.

\subsection{Supporting and Engaging Patients}
\label{findings_patient}

\subsubsection{Reducing the Distance from Me to My Clinician}
\label{opportunity_memory_load}

Although the clinician dashboard is clinician-facing, participants expressed that a CDSS could also assist patients by reducing their memory burden. Patients found that \textit{``sometimes it was hard to remember everything that happened''} (V6) during homework exercises. Patients therefore felt that having a central system that collected their self-reports and sPGD would better connect them with their clinician, facilitating better communication. V2 said that they \textit{``think all these factors could show a clinician what's going on outside of the office while you're working on everything because I mean, they can't be with you 24/7.''} V7 echoed this, expressing that the system \textit{``is really like a communication between the patient and the doctor. [...] I think it'll give [...] the doctor more valid information of what's happening during that homework assessment.''}  

\subsubsection{Achievements for the Fridge}
\label{opportunity_achievements}

Participants believed CDSS-generated visualizations of patient data (as discussed in Section \ref{opportunity_homework_review}) could convey and celebrate patient progress. Comparing a patient's responses throughout therapy is a key part of processing, asking \textit{```Do you see how these two are different? Look at that. That's so cool.' [...] And if we had pictures [...] to help with that discussion, that's also value added''} (C3). C9 recalled a \textit{``cheerleading moment''} when they received a request to print out a graph of their patient's SUDs over time \textit{``so they could put it on their refrigerator.''} Patients agreed with the value of visual indicators of progress. V7 felt that graphs were helpful because \textit{``I can see how it's actually helping me progress, and that would give me that motivation to [...] continue to do the exposure therapies.''}  V1 similarly felt that the CDSS \textit{``could potentially negate a lot of feelings of, you know, abandonment, loss, you know, [feeling] alone, things like that''} because there are more sources of support, \textit{``someone that is looking to fix it, and they're giving you the challenge of how you can fix it.''}

\subsubsection{Unpacking the Nuances of the ``PTSD Veteran'' Identity}
\label{opportunity_identity}

Most of the veterans expressed positive opinions of the CDSS, especially its potential ability to support adherence to homework exposures. V1 partially attributed this to the military identity, stating that \textit{``a lot of military people don't like to not do what they're asked to do. Because of just, you know, years of doing everything. And if you're seriously looking to fix yourself [...], I know I would do my best to make sure that happened.''}

At the same time, however, potential obstacles to full engagement were attributed to military identity. Although imaginal exposures are meant to be completed while sitting still and preventing all distractions, V3 noted that \textit{``because I came from a military background [...], instead of sitting around in a quiet place I need to do something. Running, walking, hiking [...] I don't like sitting still in a room where it’s super quiet.}'' Clinicians also described that patients \textit{``might think it’s almost disrespectful to someone who died [...] if they get better''} (C1) and discussed the ``PTSD veteran'' identity: \textit{``It's almost like, you know, `I'm a PTSD veteran, that's a part of my identity’ and `Oh, I don't want to give that up.’\thinspace ''} (C1). Veterans, especially those who have been living with PTSD for many years, may therefore face \textit{``an issue of like [...] what does it mean if I am better and I'm not feeling all of this?''} (C5) These manifestations of identity highlight the need to discuss and contextualize sPGD use tailored to veterans, and suggest that ``detective work'' (as discussed in Section \ref{opportunity_detective}) may assist clinicians in identifying this deeper and more cultural avoidance  tied to military identity.

\subsection{Contextual Challenges for Clinical Decision Support System (CDSS) Design}
\label{findings_challenges}
Our analysis also distilled three contextual challenges for a CDSS to support therapy delivery. These challenges are regarding the role the CDSS should play in patient-clinician interactions, perspectives on psychophysiology, and the complexities that arise due to the Department of Veterans Affairs (the VA).

\subsubsection{Understanding the Role of the CDSS: The ``In Between''}
\label{challenge_in_between}

Although clinicians were positive about the utility of the prototyped clinician dashboard, they were most hesitant about the digitalized session checklist page. This made salient a key concern: clinicians worried about any non-human intermediary coming ``in between'' clinician and patient. 

The concern of something coming ``in between'' is already present when clinicians use a printed out session checklist. C1 worried about seeming inattentive: \textit{``I don't like to kind of turn my back or [...] be attending to something else when there's a patient in the room.''} C1 therefore viewed the manual as \textit{``something that [...] shouldn't be open in between you and your patient [...]. The manual is not like a cookbook or intended to be, kind of, `in the room.'\thinspace''} C7 echoed this sentiment, noting that \textit{``the beauty of telehealth''} is that clinicians \textit{``can just have [the checklist] up and then no one knows.''} This covert manner stands in contrast to the PE protocol \cite{foa_prolonged_2019}, which recommends that the checklist should be used in all sessions to ensure high fidelity.  An alternative approach was to acknowledge the checklist's existence and clarify its purpose, for example by letting a patient \textit{``know that I'm using a guide to make sure I give them everything they need for that session''} (C8).

The idea of transitioning to a digital session checklist stirred negative responses from multiple clinicians. They felt that it would exacerbate the distance between clinician and patient, especially because \textit{``read[ing] the rationales off the screen [...] would mean that I was looking away from the person''} (C6). C7 similarly asserted that \textit{``especially in the context of psychotherapy, I think it's really impersonal to be staring at a screen.''} This suggests that although using a session checklist on a clipboard or an electronic dashboard are not that functionally different-- they both require looking  away from a patient briefly for informational support-- the digital aspect carries an added stigma regarding hindered interpersonal connection. This delicate spatial arrangement is further constrained by patient preferences within the clinician's office. As C2 noted, \textit{``with veterans, you know, they might pick the seat [...] that they can see the door or they can see out the window.''} A digitalized session checklist could therefore impair a clinician's ability to be flexible in the spatial arrangement of the therapy session.

\subsubsection{Reaching a Community Consensus on Psychophysiology}
\label{challenge_psychophysiology}

Almost all of the clinicians and patients believed that physical responses were indicative of emotional states, i.e., psychophysiology. When reporting their experiences of in vivo exercises, for example, patients include details such as \textit{``if my heart was racing, if I was perspiring''} (V4). V6 described the goal of the CDSS as \textit{``allow[ing] you to capture your-- I would say your emotions.''} Some also referred to sensor-captured patient-generated data (sPGD) as \textit{``an objective third party''} and \textit{``indisputable''} (C5). This strong trust in psychophysiological measures facilitated multiple perceived opportunities for sPGD. It would allow clinicians to judge \textit{``whether [patients are] accurate reporters of their internal state''} (C3) (see Section \ref{opportunity_conceptualization}) and drive discussion during clinic sessions (see Section \ref{opportunity_detective}). V1 even wondered if the data could aid differential diagnosis, \textit{``being able to use the data to determine [...] whether or not it was another mental problem other than PTSD.''} 

However, not everyone believed that physical responses were an objective representation of internal mental state. C1 was careful to note their \textit{``skepticism about the value of psychophysiology in this context. For people who haven't looked at the literature on this, it makes perfect sense that you would go and see nice reductions in psychophysiology. The literature on PTSD and psychophysiology is a mess, and it's not as clear as one would expect. There are people who don't have big physiological responding.''} Given the literature on the military population's ability to ``turn off'' emotions \cite{zwiebach2019military},  this consideration may be especially important for veteran care: \textit{``because sometimes and especially in the military population [...] they're what I call `in reporting mode.' So they're just saying what happened, but they're not really emotionally engaged''} (C4). If sPGD were incorporated into a CDSS, then, it would be vital to include educational materials for clinicians and patients to facilitate a pragmatic and conservative understanding of the role this data should play in CBT delivery, both in general and specifically for veteran care.

\subsubsection{The Elephant in the Room: Navigating Veterans Affairs}
\label{challenge_VA}

Most clinicians and patients were optimistic about the CDSS; however, Veterans Affairs (the VA) frequently arose as a key institutional challenge. As C7 put it, \textit{``the VA has a lot of, you know... barriers, I guess, is the nice word.''} Patients were aware of these ``barriers,'' too, noting \textit{``turnover of their providers''} as a \textit{``consistent problem''} (V1), and lamenting recent staffing issues: \textit{``I haven't been able to see anyone regularly''} (V2). 

Knowledge of the VA as a potential deployment context impacted clinicians' perceptions of the prototype. C1 felt skeptical about its feasibility, expressing that they \textit{``just wish that there was a way to develop it with the level of security that would allow it to be used in more systems. [...] You can look up the VA's data security requirements. They're insane. [...] So I don't think you guys will meet it, but I wish you luck.''} In contrast, C2 felt that the VA context made a CDSS (especially the quick review of patient data) more desirable because at the VA, \textit{``you might only have a couple of patients at a time who are doing PE because of the constraints on the system [...]. Like to have a 90 minute slot [...] weekly, you're not gonna have a lot of those.''}

Imagining deployment within the VA also highlights the complexities of power dynamics in play. The VA creates potential ulterior motives for patients reporting their symptoms, which may cause discrepancies between self-reports and sPGD. C5 noted that the \textit{``VA has a service connection where they will pay people money based on how symptomatic they are. So there's a lot of incentive for them to fill out questionnaires higher, right? It doesn't mean that they're not getting better.''} V1 expressed that they would feel comfortable with the collection and sharing of their data during treatment  \textit{``as long as it's not the VA [...] because [there may be situations] where you're afraid to be completely honest on certain things [...]. Like, some of my friends were afraid to say we smoke weed [...]. Other people are afraid if they say [...] `I'm feeling better,' will it take away from their percentages of the money they receive?''}  Although the clinician dashboard may appear to be a simple informational decision support system, this technology is clearly not power-neutral. When deployed within a real CBT context, especially within the VA, a CDSS would collect patient data that could enhance current institutional power dynamics.

\section{Discussion} 
\label{discussion_alternative_views}

Previous studies in clinical decision support system (CDSS) literature \cite{jo_designing_2022,lee_towards_2022,lee_human-ai_2021,yang_unremarkable_2019,yang_investigating_2016} highlight the importance of contextual fit and a user-centered approach for successful design. Because of the mental health workforce shortage \cite{blane2013cognitive,oconnorIncreasingAvailabilityPsychological2018} and high-risk clinical contexts, any technology intervention must not disrupt existing workflows. In highly restrictive and sensitive clinical settings such as psychotherapy, however, it is difficult to follow the typical design process \cite{yang_investigating_2016}. 

We conducted an interview study with 9 clinicians and 7 former patients (and veterans of the United States military), to deepen our understanding of Prolonged Exposure (PE) therapy within the context of intensive outpatient veteran care, and explore their perspectives on a CDSS. As a technology probe to assist clinician interviews, we designed a Figma prototype of a clinician dashboard to support PE therapy delivery. Because the sensitive nature of psychotherapy prevents researchers from direct observation, initial prototype design was based on prior related work, the PE manual \cite{foa_prolonged_2019}, and the perspectives of three PE experts (including our clinical co-author). Our findings revealed a number of opportunities for a CDSS, as well as three contextual challenges for deployment. 

We now reframe our findings to explore three human-centered lenses on the sociotechnical system of PE therapy for veteran care. These perspectives were intentionally chosen through team discussion because they cover all of the themes distilled from our thematic analysis, yet their different focuses result in varying (or even contradictory) design guidelines. In Section \ref{discussion_dicog}, we leverage distributed cognition theory to discuss informational flow and spatial, temporal, and social distribution within PE therapy. Moving away from information flow, we use situated learning in Section \ref{discussion_learning} to question a CDSS's impact on patient learning and power dynamics. Finally, in Section \ref{discussion_infrastructure} we discuss the PE protocol as a key aspect of PE infrastructure and explore how introducing a CDSS into the sociotechnical system might incite infrastructural inversion. For each perspective, we offer design considerations and avenues for future work for the design or analysis of a CDSS within this context.

\subsection{A Distributed Cognition Perspective: Information Flow and Cognitive Burden}
\label{discussion_dicog}

PE therapy delivery can be viewed as a complex sociotechnical system with the ultimate goal of helping patients process their trauma and reduce trauma-related symptoms. This process is mediated by many internal and external cognitive representations that facilitate information flow between social actors and artifacts over time. Designers should therefore consider how a CDSS's impacts on the social, spatial, and temporal distribution \cite{Hollan2000} of information influence the overall collaborative goal of successful therapy delivery. Below we reframe some of our findings through these spatially, temporally, and socially shared cognitive representations. 

Clinicians rely on a variety of physical and electronic tools, e.g., manuals, checklists, notes, graphs. Clinicians \textbf{spatially} distribute to these artifacts rather than relying only on their memory. Our findings suggest that a clinician dashboard could act as a long-term memory aid by integrating heterogeneous clinician records in one place (Section \ref{opportunity_homework_review}). Flexibility, however, would be crucial for this new external representation. The extent to which clinicians spatially distribute varies, with less experienced clinicians desiring more informational support (Section \ref{opportunity_info_support}). To accommodate differing needs, future CDSSs should allow clinicians to personalize this level of support. This may avoid adoption resistance stemming from a feeling that the CDSS is undermining their workflow \cite{jo_designing_2022,khairat_reasons_2018,wang_brilliant_2021,yang_investigating_2016}. The CDSS should also provide flexibility on which data streams are included and how, especially until a community consensus is reached on psychophysiology (Section \ref{challenge_psychophysiology}). In the mean time, treating sensor-captured patient-generated data (sPGD) as augmenting rather than replacing self-report measures could increase system redundancy, which is an essential component of a safe sociotechnical system \cite{Hutchins1995}. Still, an open question is to what extent a CDSS should enable personalization to support an individual clinician's approach and perspective, as opposed to remaining accountable only to the clinical protocol. Similarly, for patients, tailoring the data streams collected by the CDSS (and enabling review and deletion of any data streams) could allow for a variety of avenues to ``opt in'' to  monitoring. While the former patients we interviewed expressed that they were comfortable sharing  data with their clinicians (Section \ref{challenge_VA}), prior work has highlighted the importance of clear guidelines around data privacy and storage for sPGD, especially for military populations \cite{evansUsingSensorCapturedPatientGenerated2024,evans_understanding_2020}. 

PE therapy is a longitudinal process ranging from 2 to 15 weeks \cite{foa_prolonged_2019}, but understanding patient progress relies on holistic interpretation in partnership with a clinician. \textbf{Temporal} and \textbf{social} distribution are crucial, especially as a patient's memory of their homework exposures will naturally decay. Our findings indicate that a CDSS could enable quick review of patient homework (Section \ref{opportunity_homework_review}) and reduce burdens on patient memory (Section \ref{opportunity_memory_load}). In particular, sPGD may act as a proxy representing some of the patient’s experience, cueing to narrow temporal gaps. The passive nature of sPGD may mitigate the tracking burden of patients  \cite{chung_finding_2017,li2010stage,oh_patients_2022}. Although interpreting and discussing patient homework is integral to PE therapy, introducing new heterogeneous data streams could also alter the clinical workflow. Clinicians might take extra time to review sPGD before sessions, as well as time during sessions to assist in reframing patients' interpretations of their data \cite{ng2019provider,west_common_2018}. Researchers should additionally consider how enabling the distribution of patient cognition through sPGD could impact a patient's emotional engagement and processing, especially centering the nuances of emotional engagement for veterans who enter ``reporting mode'' (C4) (Section \ref{challenge_psychophysiology}). 

Although patient and clinician are the most visible actors in the PE system, others are involved in \textbf{social} distribution. Access to a patient's homework data  could facilitate discussion between newer clinicians and their supervisors (Section \ref{opportunity_homework_review}). Future design may consider including functionality to support this use case, for example by allowing annotation \cite{mentis_crafting_2017} and marking of specific timestamps or data streams. Our findings also suggest the opportunity for a CDSS to aid organizational transfers, both unexpected (e.g., clinician absences due to illness) and expected (i.e., throughout a patient's mental health journey) (Section \ref{opportunity_transfers}). Many CBTs, including intensive outpatient PE, feature assessments and interventions by a multidisciplinary clinical team \cite{howard2023cognitive,rauch2020prolonged}.  Extending our findings, future research could explicitly explore design opportunities to support this team-based approach, and question how a CDSS would impact the equity of cognitive burden throughout the sociotechnical system, e.g., patient, clinician, clinical supervisor, other clinicians involved in team-based care.

\subsection{A Situated Learning Perspective: Power and Access}
\label{discussion_learning}

Psychotherapy can be seen as a learning environment \cite{bandura1961psychotherapy,rose2005counselling}, in which a learner (patient) and teacher (clinician) collaboratively work towards patient learning (e.g., that reminders of trauma are safe). Throughout this learning, the patient's perspectives shift and they learn the shared language of the therapy. An emphasis on in vivo homework exercises to encourage learning in real-world contexts aligns with the theory of  situated learning \cite{lave1991situated}, which views learning as inseparable from context. 

Our findings indicate that our prototyped dashboard could aid a clinician's conceptualization of their patient while learning through in vivo exposures in the real world (Section \ref{opportunity_conceptualization}). These conceptualizations could provide clinicians additional contextual clues regarding a patient's homework, including discrepancies that could hinder situated learning by reducing the similarity of the homework context to real-world contexts (e.g., if Spotify was on during the exposure). In other words, the CDSS data might provide ways for clinicians to better ensure and enforce that patient learning is occurring within true real-world contexts. Despite literature that suggests sPGD would empower patients \cite{evansUsingSensorCapturedPatientGenerated2024,ng2019provider,khatiwada2024patient}, this raises questions regarding patient agency. Patients would lose some control if  patient conceptualization bypasses the patient, i.e., when the CDSS and sPGD are leveraged. Military culture's emphasis on obedience and hierarchy \cite{collins1998complex} raises questions of data control and privacy: As V1 stated, \textit{ ``a lot of military people don’t like to not do what they’re asked to do''} (Section \ref{opportunity_identity}). Exploring cultural implications for willingness to share sPGD could uncover avenues to bolster informed consent procedures for veterans undergoing psychotherapy. It is also unknown whether (and if so, how) introducing additional quantitative data streams representing patient experience would impact the ``Veteran PTSD'' identity (Section \ref{opportunity_identity}) and perspective shifts that these learners undergo during therapy.

Situated learning as an analytical perspective may also help understand power dynamics within a learning environment, as a CDSS could both facilitate and hinder patient access to learning opportunities and empowerment within their treatment. In one respect, the automatic creation of visualizations that show evidence of patient progress (Section \ref{opportunity_achievements}) could be useful to a patient if available to them-- evidence that was once only available to the clinician. These indications of progress could be especially useful in veteran care, as previous work exploring veteran dropout of trauma-focused treatments notes the importance of encouragement and believing that the therapy is working \cite{hundt2020didn,kehle2022divergent}. At the same time, however, the additional data within the CDSS, especially sPGD, could enable clinicians to do more ``detective'' work regarding a patient's experience (Section \ref{opportunity_detective}). While this better patient conceptualization could facilitate a more learner-centered environment, it would also fundamentally shift the power dynamics because patients previously had complete control (within the limitations of their distress) over what and how they self-reported. One could imagine an alternative patient-facing system that offers the learning opportunities created by this ``detective'' work to patients rather than clinicians.

\subsection{An Infrastructural Inversion Perspective: Consensus and (In)Visibility} 
\label{discussion_infrastructure}

Star and Ruhleder described the properties of infrastructure, including its extensive reach, that it is learned through membership, and becomes visible only upon breakdown \cite{star1996steps}. The clinical manual is a key aspect of infrastructure within the sociotechnical system of PE therapy. The associated protocol is learned by all PE clinicians, dictates PE therapy practice, and yet is largely taken for granted once internalized. This invisibility was underlined by C1's comment that the manual is not ``intended to be [...] `in the room.'\thinspace'' A CDSS, once successfully deployed and adopted, could  become an aspect of infrastructure as well, for example facilitating more seamless transfers of patient data between clinicians (Section \ref{opportunity_transfers}). At the same time, our findings suggest that introducing a CDSS within this context in the first place could facilitate infrastructural inversion, making the typically invisible more visible. Digitalizing the clinical manual into a CDSS could bring the manual into the foreground, driving clinician concerns that it would come ``in between'' clinician and patient (Section \ref{challenge_in_between}). How would increased awareness and presence of the clinical protocol as ``in the room'' change patient-clinician collaboration and patient perspectives on the therapy? We question if this could potentially be empowering for patients and level the playing field, facilitating collaboration. So much is implied by the clinical protocol, including epistemologies and assumptions of patients. While patients consent to PE therapy due to its high clinical efficacy, persistent critiques of biomedicalization and psychiatric diagnoses \cite{frieh2024resistance,karter2025ecological} throw into question some of the assumptions within clinical protocols. Bringing this to the foreground could bring more former PE patients into discussions around desired changes to the therapeutic approach. For example, collaboration between clinicians and veterans who are former PE patients could generate avenues to better support patients struggling with their identity as a ``PTSD veteran'' (Section \ref{opportunity_identity}).

In addition, a CDSS could make more apparent the heterogeneity of the clinical community, highlighting current issues or disagreements. For example, wide deployment of CDSSs like our prototyped dashboard could spotlight debates regarding psychophysiology. While many clinicians and their patients rely on physical indications of emotional state, others note the complexities and knowledge gaps in the literature, especially for military populations (Section \ref{challenge_psychophysiology}). This may be an aspect of clinician positionality that is typically not reflected upon nor discussed with patients, even though it has implications for treatment and interpretation of patient progress. Highlighting this discrepancy and interrogating this lack of consensus may ultimately forge a path towards more effective technology to support CBT, for example through utilizing sPGD if deemed appropriate. It may be even more relevant to this specific context of veteran care, as veterans' military identity can contribute to a more numb and calm presentation \cite{zwiebach2019military}, decreasing physiological indicators of distress. The question of psychophysiology also positions novel technology as a catalyst for raising additional questions: What other lacks of consensus could the deployment of a CDSS within CBT uncover? Designers should reflect on their responsibility in accommodating multiple perspectives on debated topics (such as the validity of psychophysiology) if they arise during the design process.

An additional aspect of critical infrastructure emerged during interviews: Veterans Affairs. Due to the power dynamics involved, the VA creates powerful ulterior motives for patient self-reports of symptoms (Section \ref{challenge_VA}). Because a CDSS could collect patient data aligning with the institutional power of the VA, encompassing the largest integrated health care system in the country \cite{VHAStats}, it could make the VA even more visible during treatment. The clinicians and former patients we interviewed mentioned many ``barriers'' (C7) of the VA that were at the forefront of their minds, including C1's skepticism that our research team would ever be able to meet their security standards (Section \ref{challenge_VA}). Although the VA was highly influential for most of our participants, the VA was perhaps most salient in our study design when planning and conducting participant recruitment: Many of our clinician participants' appointments with the VA facilitated easier snowball sampling. Our team's consideration of the VA was only cursory in the design of our prototype. After all, infrastructure is invisible only to those who do not experience its breakdowns, and those who are not tasked with doing the continuous work of maintenance and upkeep. 

\section{Limitations \& Future Work}

The nine PE clinicians we interviewed had varying years of experience; however, they all had doctorate degrees and worked at nationally recognized programs that conduct research about PE therapy and technological tools to support delivery. These clinicians may therefore have high receptiveness to the use of novel technology in their practice. Clinicians from other settings (e.g., clinicians from community-based organizations or those practicing PE within primary care settings \cite{rauch2023treatment}) might have different needs and perceptions of new technology. Future studies should explore technological support for clinical practice conducted by mental health professionals with more diverse educational backgrounds, such as those with master’s degrees (e.g., in social work or family and child counseling), and who practice in different settings. 

We also interviewed seven former PE patients, which is unlikely to capture the diversity of patient perspectives on incorporating technology into PTSD treatment. In fact, all of the former patients we interviewed were veterans of the U.S. military, recruited by our clinical co-author from a veterans' intensive outpatient program. By deploying a CDSS with an additional patient-facing interface, future researchers could investigate to what extent the system supports patient experiences in the real world. This would allow for a more ecologically valid examination of the impacts of the system, such as effects on clinical conversations, and how sensor data is used in practice. In addition, future work could explore opportunities to support standard PE therapy provided by the VA.

All of the former PE patients we interviewed were optimistic about a CDSS supporting PE therapy. There is a broad body of literature, however, exploring the ethical implications of sPGD and passive tracking of health data (e.g.,  \cite{evans_understanding_2020,evansUsingSensorCapturedPatientGenerated2024,khatiwada2024patient,hogan2025veteran,saka2025watch,chung_boundary_2016}). Future work should explicitly center factors including patient privacy, data security, and trust, to distill further design implications for a CDSS for veteran PTSD treatment. 

This study explored only a part of the sociotechnical system of manualized psychotherapy to treat veterans with PTSD. Our work centered clinicians and former patients, identifying opportunities for technology support via a CDSS. Previous literature has noted, however, that leveraging social networks (e.g., family and friends) can support the data practice of patients with various mental health conditions \cite{evans_perspectives_2022,foong_harvesting_2018,hong_investigating_2013,jacobs_comparing_2015,murnane_personal_2018,schertz_bridging_2019,evans_understanding_2020}. Evans et al. \cite{evans_perspectives_2022} examined integrating the perspectives from trusted others into PE therapy, but there is still a knowledge gap regarding the benefits and challenges of integrating this additional viewpoint to better understand patient experience and therapy progress. Future research may explore these new dynamics to better understand the situated objectivity \cite{ng2019provider, pantzar_living_2017} of patient experience.

Finally, our findings drove an exploration of three human-centered perspectives on intensive outpatient PTSD veteran care as a sociotechnical system: distributed cognition, situated learning, and infrastructural inversion. These perspectives are by no means comprehensive, and we look forward to future research that reveals the utility of additional lenses for better understanding and designing for psychotherapeutic contexts.
\section{Conclusion}

Psychotherapy practice relies heavily on a negotiation between patient self-reports and clinical intuition. To explore the ways in which technology may be able to aid clinicians in mediating this tension, we designed a prototype of a clinical decision support system (CDSS) to support the delivery of psychotherapy for veterans with PTSD in an intensive outpatient setting. We conducted a two-phase interview study with practicing PE clinicians and former PE patients (who were all United States veterans) and distilled opportunities and challenges for a CDSS in this context. Then, by reframing our findings through three human-centered perspectives (distributed cognition, situated learning, infrastructural inversion), we offer various theory-aligned design considerations and avenues for future work. Our work highlights the complexities of building a CDSS for veteran PTSD care, which can be viewed simultaneously as a goal-oriented and collaborative system, as a learning environment emphasizing real-world learning, and as a space laden with organizational and institutional power dynamics.

\begin{acks}
We would like to thank Hayley Evans and Jiawei Zhou for assistance with data collection. This work was supported by the National Science Foundation under Award Number 1915504.
\end{acks}

\bibliographystyle{ACM-Reference-Format}
\bibliography{_references}


\begin{thebibliography}{156}


\ifx \showCODEN    \undefined \def \showCODEN     #1{\unskip}     \fi
\ifx \showISBNx    \undefined \def \showISBNx     #1{\unskip}     \fi
\ifx \showISBNxiii \undefined \def \showISBNxiii  #1{\unskip}     \fi
\ifx \showISSN     \undefined \def \showISSN      #1{\unskip}     \fi
\ifx \showLCCN     \undefined \def \showLCCN      #1{\unskip}     \fi
\ifx \shownote     \undefined \def \shownote      #1{#1}          \fi
\ifx \showarticletitle \undefined \def \showarticletitle #1{#1}   \fi
\ifx \showURL      \undefined \def \showURL       {\relax}        \fi
\providecommand\bibfield[2]{#2}
\providecommand\bibinfo[2]{#2}
\providecommand\natexlab[1]{#1}
\providecommand\showeprint[2][]{arXiv:#2}

\bibitem[Addis and Krasnow(2000)]%
        {addis2000national}
\bibfield{author}{\bibinfo{person}{Michael~E Addis} {and} \bibinfo{person}{Aaron~D Krasnow}.} \bibinfo{year}{2000}\natexlab{}.
\newblock \showarticletitle{A national survey of practicing psychologists' attitudes toward psychotherapy treatment manuals.}
\newblock \bibinfo{journal}{\emph{Journal of consulting and clinical psychology}} \bibinfo{volume}{68}, \bibinfo{number}{2} (\bibinfo{year}{2000}), \bibinfo{pages}{331}.
\newblock


\bibitem[Aguilera and Muench(2014)]%
        {aguilera_theres_2014}
\bibfield{author}{\bibinfo{person}{Adrian Aguilera} {and} \bibinfo{person}{Frederick Muench}.} \bibinfo{year}{2014}\natexlab{}.
\newblock \showarticletitle{There's an {App} for {That}: {Information} {Technology} {Applications} for {Cognitive} {Behavioral} {Practitioners}}.
\newblock \bibinfo{journal}{\emph{Behav Ther}} (\bibinfo{year}{2014}).
\newblock


\bibitem[Are{\'a}n et~al\mbox{.}(2021)]%
        {arean2021using}
\bibfield{author}{\bibinfo{person}{Patricia~A Are{\'a}n}, \bibinfo{person}{Emily~C Friedman}, \bibinfo{person}{Abhishek Pratap}, \bibinfo{person}{Ryan Allred}, \bibinfo{person}{Jaden Duffy}, \bibinfo{person}{Sara Gille}, \bibinfo{person}{Shelley Reetz}, \bibinfo{person}{Erin Keast}, {and} \bibinfo{person}{Gregory Clarke}.} \bibinfo{year}{2021}\natexlab{}.
\newblock \showarticletitle{Using Real-world Data for Decision Support: Recommendations from a Primary Care Provider Survey}.
\newblock \bibinfo{journal}{\emph{The Permanente Journal}}  \bibinfo{volume}{25} (\bibinfo{year}{2021}).
\newblock


\bibitem[Association(2013)]%
        {american2013diagnostic}
\bibfield{author}{\bibinfo{person}{American~Psychiatric Association}.} \bibinfo{year}{2013}\natexlab{}.
\newblock \bibinfo{booktitle}{\emph{Diagnostic and statistical manual of mental disorders: DSM-5}}. Vol.~\bibinfo{volume}{5}.
\newblock


\bibitem[Bandura(1961)]%
        {bandura1961psychotherapy}
\bibfield{author}{\bibinfo{person}{Albert Bandura}.} \bibinfo{year}{1961}\natexlab{}.
\newblock \showarticletitle{Psychotherapy as a learning process.}
\newblock \bibinfo{journal}{\emph{Psychological Bulletin}} \bibinfo{volume}{58}, \bibinfo{number}{2} (\bibinfo{year}{1961}).
\newblock


\bibitem[Bardram and Frost(2013)]%
        {bardram_designing_2013}
\bibfield{author}{\bibinfo{person}{Jakob~E Bardram} {and} \bibinfo{person}{Mads Frost}.} \bibinfo{year}{2013}\natexlab{}.
\newblock \showarticletitle{Designing mobile health technology for bipolar disorder: a field trial of the monarca system}.
\newblock \bibinfo{journal}{\emph{Mental Health}} (\bibinfo{year}{2013}), \bibinfo{pages}{10}.
\newblock


\bibitem[Bardram et~al\mbox{.}(2016)]%
        {bardram_designing_2016}
\bibfield{author}{\bibinfo{person}{Jakob~E Bardram}, \bibinfo{person}{Mads Frost}, \bibinfo{person}{Nanna Tuxen}, \bibinfo{person}{Maria Faurholt-Jepsen}, {and} \bibinfo{person}{Lars~V Kessing}.} \bibinfo{year}{2016}\natexlab{}.
\newblock \showarticletitle{Designing context-aware cognitive behavioral therapy for unipolar and bipolar disorders}. In \bibinfo{booktitle}{\emph{Proceedings of the 2016 {ACM} {International} {Joint} {Conference} on {Pervasive} and {Ubiquitous} {Computing}: {Adjunct}}}. \bibinfo{publisher}{ACM}, \bibinfo{pages}{1162--1170}.
\newblock
\showISBNx{978-1-4503-4462-3}


\bibitem[Bardram and Matic(2020)]%
        {bardram_decade_2020}
\bibfield{author}{\bibinfo{person}{Jakob~E Bardram} {and} \bibinfo{person}{Aleksandar Matic}.} \bibinfo{year}{2020}\natexlab{}.
\newblock \showarticletitle{A {Decade} of {Ubiquitous} {Computing} {Research} in {Mental} {Health}}.
\newblock \bibinfo{journal}{\emph{IEEE Pervasive Computing}} \bibinfo{volume}{19}, \bibinfo{number}{1} (\bibinfo{date}{Jan.} \bibinfo{year}{2020}), \bibinfo{pages}{62--72}.
\newblock
\showISSN{1536-1268, 1558-2590}
\href{https://doi.org/10.1109/MPRV.2019.2925338}{doi:\nolinkurl{10.1109/MPRV.2019.2925338}}


\bibitem[Barnett(2019)]%
        {barnettEthicalPracticePsychotherapy2019}
\bibfield{author}{\bibinfo{person}{Jeffrey~E Barnett}.} \bibinfo{year}{2019}\natexlab{}.
\newblock \showarticletitle{The ethical practice of psychotherapy: {Clearly} within our reach.}
\newblock \bibinfo{journal}{\emph{Psychotherapy}} \bibinfo{volume}{56}, \bibinfo{number}{4} (\bibinfo{year}{2019}).
\newblock
\showISSN{1939-1536, 0033-3204}


\bibitem[Beck and Alford(2009)]%
        {beck2009depression}
\bibfield{author}{\bibinfo{person}{Aaron~T Beck} {and} \bibinfo{person}{Brad~A Alford}.} \bibinfo{year}{2009}\natexlab{}.
\newblock \bibinfo{booktitle}{\emph{Depression: Causes and treatment}}.
\newblock \bibinfo{publisher}{University of Pennsylvania Press}.
\newblock


\bibitem[Beede et~al\mbox{.}(2020)]%
        {beede_human-centered_2020}
\bibfield{author}{\bibinfo{person}{Emma Beede}, \bibinfo{person}{Elizabeth Baylor}, \bibinfo{person}{Fred Hersch}, \bibinfo{person}{Anna Iurchenko}, \bibinfo{person}{Lauren Wilcox}, \bibinfo{person}{Paisan Ruamviboonsuk}, {and} \bibinfo{person}{Laura~M. Vardoulakis}.} \bibinfo{year}{2020}\natexlab{}.
\newblock \showarticletitle{A {Human}-{Centered} {Evaluation} of a {Deep} {Learning} {System} {Deployed} in {Clinics} for the {Detection} of {Diabetic} {Retinopathy}}. In \bibinfo{booktitle}{\emph{Proceedings of the 2020 {CHI} {Conference} on {Human} {Factors} in {Computing} {Systems}}}. \bibinfo{publisher}{ACM}, \bibinfo{pages}{1--12}.
\newblock
\showISBNx{978-1-4503-6708-0}
\href{https://doi.org/10.1145/3313831.3376718}{doi:\nolinkurl{10.1145/3313831.3376718}}


\bibitem[Bhattacharya et~al\mbox{.}(2023)]%
        {bhattacharya2023directive}
\bibfield{author}{\bibinfo{person}{Aditya Bhattacharya}, \bibinfo{person}{Jeroen Ooge}, \bibinfo{person}{Gregor Stiglic}, {and} \bibinfo{person}{Katrien Verbert}.} \bibinfo{year}{2023}\natexlab{}.
\newblock \showarticletitle{Directive explanations for monitoring the risk of diabetes onset: introducing directive data-centric explanations and combinations to support what-if explorations}. In \bibinfo{booktitle}{\emph{Proceedings of the 28th international conference on intelligent user interfaces}}. \bibinfo{pages}{204--219}.
\newblock


\bibitem[Bisson et~al\mbox{.}(2013)]%
        {bisson2013psychological}
\bibfield{author}{\bibinfo{person}{Jonathan~I Bisson}, \bibinfo{person}{Neil~P Roberts}, \bibinfo{person}{Martin Andrew}, \bibinfo{person}{Rosalind Cooper}, {and} \bibinfo{person}{Catrin Lewis}.} \bibinfo{year}{2013}\natexlab{}.
\newblock \showarticletitle{Psychological therapies for chronic post-traumatic stress disorder (PTSD) in adults}.
\newblock \bibinfo{journal}{\emph{Cochrane database of systematic reviews}} \bibinfo{number}{12} (\bibinfo{year}{2013}).
\newblock


\bibitem[Blandford and Furniss(2005)]%
        {Blandford2005}
\bibfield{author}{\bibinfo{person}{Ann Blandford} {and} \bibinfo{person}{Dominic Furniss}.} \bibinfo{year}{2005}\natexlab{}.
\newblock \showarticletitle{DiCoT: a methodology for applying distributed cognition to the design of teamworking systems}. In \bibinfo{booktitle}{\emph{International workshop on design, specification, and verification of interactive systems}}. Springer.
\newblock


\bibitem[Blane et~al\mbox{.}(2013)]%
        {blane2013cognitive}
\bibfield{author}{\bibinfo{person}{David Blane}, \bibinfo{person}{Chris Williams}, \bibinfo{person}{Jill Morrison}, \bibinfo{person}{Alistair Wilson}, {and} \bibinfo{person}{Stewart Mercer}.} \bibinfo{year}{2013}\natexlab{}.
\newblock \showarticletitle{Cognitive behavioural therapy: why primary care should have it all}.
\newblock \bibinfo{journal}{\emph{British Journal of General Practice}} \bibinfo{volume}{63}, \bibinfo{number}{607} (\bibinfo{year}{2013}), \bibinfo{pages}{103--104}.
\newblock


\bibitem[Blevins et~al\mbox{.}(2015)]%
        {blevins2015posttraumatic}
\bibfield{author}{\bibinfo{person}{Christy~A Blevins}, \bibinfo{person}{Frank~W Weathers}, \bibinfo{person}{Margaret~T Davis}, \bibinfo{person}{Tracy~K Witte}, {and} \bibinfo{person}{Jessica~L Domino}.} \bibinfo{year}{2015}\natexlab{}.
\newblock \showarticletitle{The posttraumatic stress disorder checklist for DSM-5 (PCL-5): Development and initial psychometric evaluation}.
\newblock \bibinfo{journal}{\emph{Journal of traumatic stress}} \bibinfo{volume}{28}, \bibinfo{number}{6} (\bibinfo{year}{2015}), \bibinfo{pages}{489--498}.
\newblock


\bibitem[Bower and Gilbody(2005)]%
        {bower2005stepped}
\bibfield{author}{\bibinfo{person}{Peter Bower} {and} \bibinfo{person}{Simon Gilbody}.} \bibinfo{year}{2005}\natexlab{}.
\newblock \showarticletitle{Stepped care in psychological therapies: access, effectiveness and efficiency: narrative literature review}.
\newblock \bibinfo{journal}{\emph{The British Journal of Psychiatry}} \bibinfo{volume}{186}, \bibinfo{number}{1} (\bibinfo{year}{2005}), \bibinfo{pages}{11--17}.
\newblock


\bibitem[Bowker(1994)]%
        {bowker1994science}
\bibfield{author}{\bibinfo{person}{Geoffrey~C Bowker}.} \bibinfo{year}{1994}\natexlab{}.
\newblock \bibinfo{booktitle}{\emph{Science on the run: Information management and industrial geophysics at Schlumberger, 1920-1940}}.
\newblock \bibinfo{publisher}{MIT press}.
\newblock


\bibitem[Bowker and Star(2000)]%
        {bowker2000sorting}
\bibfield{author}{\bibinfo{person}{Geoffrey~C Bowker} {and} \bibinfo{person}{Susan~Leigh Star}.} \bibinfo{year}{2000}\natexlab{}.
\newblock \bibinfo{booktitle}{\emph{Sorting things out: Classification and its consequences}}.
\newblock \bibinfo{publisher}{MIT press}.
\newblock


\bibitem[Braun and Clarke(2006)]%
        {braun2006using}
\bibfield{author}{\bibinfo{person}{Virginia Braun} {and} \bibinfo{person}{Victoria Clarke}.} \bibinfo{year}{2006}\natexlab{}.
\newblock \showarticletitle{Using thematic analysis in psychology}.
\newblock \bibinfo{journal}{\emph{Qualitative research in psychology}} \bibinfo{volume}{3}, \bibinfo{number}{2} (\bibinfo{year}{2006}), \bibinfo{pages}{77--101}.
\newblock


\bibitem[Braun and Clarke(2019)]%
        {braun2019reflecting}
\bibfield{author}{\bibinfo{person}{Virginia Braun} {and} \bibinfo{person}{Victoria Clarke}.} \bibinfo{year}{2019}\natexlab{}.
\newblock \showarticletitle{Reflecting on reflexive thematic analysis}.
\newblock \bibinfo{journal}{\emph{Qualitative research in sport, exercise and health}} \bibinfo{volume}{11}, \bibinfo{number}{4} (\bibinfo{year}{2019}), \bibinfo{pages}{589--597}.
\newblock


\bibitem[Braund et~al\mbox{.}(2022)]%
        {braund_smartphone_2022}
\bibfield{author}{\bibinfo{person}{Taylor~A Braund}, \bibinfo{person}{May~The Zin}, \bibinfo{person}{Tjeerd~W Boonstra}, \bibinfo{person}{Quincy J~J Wong}, \bibinfo{person}{Mark~E Larsen}, \bibinfo{person}{Helen Christensen}, \bibinfo{person}{Gabriel Tillman}, {and} \bibinfo{person}{Bridianne O’Dea}.} \bibinfo{year}{2022}\natexlab{}.
\newblock \showarticletitle{Smartphone {Sensor} {Data} for {Identifying} and {Monitoring} {Symptoms} of {Mood} {Disorders}: {A} {Longitudinal} {Observational} {Study}}.
\newblock \bibinfo{journal}{\emph{JMIR Mental Health}} \bibinfo{volume}{9}, \bibinfo{number}{5} (\bibinfo{date}{May} \bibinfo{year}{2022}).
\newblock
\showISSN{2368-7959}
\href{https://doi.org/10.2196/35549}{doi:\nolinkurl{10.2196/35549}}


\bibitem[Cai et~al\mbox{.}(2019)]%
        {cai_hello_2019}
\bibfield{author}{\bibinfo{person}{Carrie~J. Cai}, \bibinfo{person}{Samantha Winter}, \bibinfo{person}{David Steiner}, \bibinfo{person}{Lauren Wilcox}, {and} \bibinfo{person}{Michael Terry}.} \bibinfo{year}{2019}\natexlab{}.
\newblock \showarticletitle{"{Hello} {AI}": {Uncovering} the {Onboarding} {Needs} of {Medical} {Practitioners} for {Human}-{AI} {Collaborative} {Decision}-{Making}}.
\newblock \bibinfo{journal}{\emph{Proceedings of the ACM on Human-Computer Interaction}} \bibinfo{volume}{3}, \bibinfo{number}{CSCW} (\bibinfo{date}{Nov.} \bibinfo{year}{2019}), \bibinfo{pages}{1--24}.
\newblock
\showISSN{2573-0142}
\href{https://doi.org/10.1145/3359206}{doi:\nolinkurl{10.1145/3359206}}


\bibitem[Chow et~al\mbox{.}(2024)]%
        {chow_feeling_2024}
\bibfield{author}{\bibinfo{person}{Kevin Chow}, \bibinfo{person}{Thomas Fritz}, \bibinfo{person}{Liisa Holsti}, \bibinfo{person}{Skye Barbic}, {and} \bibinfo{person}{Joanna McGrenere}.} \bibinfo{year}{2024}\natexlab{}.
\newblock \showarticletitle{Feeling {Stressed} and {Unproductive}? {A} {Field} {Evaluation} of a {Therapy}-{Inspired} {Digital} {Intervention} for {Knowledge} {Workers}}.
\newblock \bibinfo{journal}{\emph{ACM Transactions on Computer-Human Interaction}} \bibinfo{volume}{31}, \bibinfo{number}{1} (\bibinfo{date}{Feb.} \bibinfo{year}{2024}), \bibinfo{pages}{1--33}.
\newblock
\showISSN{1073-0516, 1557-7325}
\href{https://doi.org/10.1145/3609330}{doi:\nolinkurl{10.1145/3609330}}


\bibitem[Chung et~al\mbox{.}(2016)]%
        {chung_boundary_2016}
\bibfield{author}{\bibinfo{person}{Chia-Fang Chung}, \bibinfo{person}{Kristin Dew}, \bibinfo{person}{Allison Cole}, \bibinfo{person}{Jasmine Zia}, \bibinfo{person}{James Fogarty}, \bibinfo{person}{Julie~A. Kientz}, {and} \bibinfo{person}{Sean~A. Munson}.} \bibinfo{year}{2016}\natexlab{}.
\newblock \showarticletitle{Boundary {Negotiating} {Artifacts} in {Personal} {Informatics}: {Patient}-{Provider} {Collaboration} with {Patient}-{Generated} {Data}}. In \bibinfo{booktitle}{\emph{Proceedings of the 19th {ACM} {Conference} on {Computer}-{Supported} {Cooperative} {Work} \& {Social} {Computing}}}. \bibinfo{publisher}{ACM}, \bibinfo{address}{San Francisco California USA}, \bibinfo{pages}{770--786}.
\newblock
\showISBNx{978-1-4503-3592-8}
\href{https://doi.org/10.1145/2818048.2819926}{doi:\nolinkurl{10.1145/2818048.2819926}}


\bibitem[Chung et~al\mbox{.}(2017)]%
        {chung_finding_2017}
\bibfield{author}{\bibinfo{person}{Chia-Fang Chung}, \bibinfo{person}{Nanna Gorm}, \bibinfo{person}{Irina~A. Shklovski}, {and} \bibinfo{person}{Sean Munson}.} \bibinfo{year}{2017}\natexlab{}.
\newblock \showarticletitle{Finding the {Right} {Fit}: {Understanding} {Health} {Tracking} in {Workplace} {Wellness} {Programs}}. In \bibinfo{booktitle}{\emph{Proceedings of the 2017 {CHI} {Conference} on {Human} {Factors} in {Computing} {Systems}}}. \bibinfo{publisher}{ACM}.
\newblock
\showISBNx{978-1-4503-4655-9}
\href{https://doi.org/10.1145/3025453.3025510}{doi:\nolinkurl{10.1145/3025453.3025510}}


\bibitem[Cohen et~al\mbox{.}(2010)]%
        {Cohen2010}
\bibfield{author}{\bibinfo{person}{Beth~E Cohen}, \bibinfo{person}{Kris Gima}, \bibinfo{person}{Daniel Bertenthal}, \bibinfo{person}{Sue Kim}, \bibinfo{person}{Charles~R Marmar}, {and} \bibinfo{person}{Karen~H Seal}.} \bibinfo{year}{2010}\natexlab{}.
\newblock \showarticletitle{Mental health diagnoses and utilization of VA non-mental health medical services among returning Iraq and Afghanistan veterans}.
\newblock \bibinfo{journal}{\emph{Journal of general internal medicine}}  \bibinfo{volume}{25} (\bibinfo{year}{2010}), \bibinfo{pages}{18--24}.
\newblock


\bibitem[Collins(1998)]%
        {collins1998complex}
\bibfield{author}{\bibinfo{person}{Joseph~J Collins}.} \bibinfo{year}{1998}\natexlab{}.
\newblock \showarticletitle{The complex context of American military culture: A practitioner's view}.
\newblock \bibinfo{journal}{\emph{Washington Quarterly}} \bibinfo{volume}{21}, \bibinfo{number}{4} (\bibinfo{year}{1998}), \bibinfo{pages}{213--228}.
\newblock


\bibitem[Coyle et~al\mbox{.}(2007)]%
        {coyleComputersTalkbasedMental2007}
\bibfield{author}{\bibinfo{person}{David Coyle}, \bibinfo{person}{Gavin Doherty}, \bibinfo{person}{Mark Matthews}, {and} \bibinfo{person}{John Sharry}.} \bibinfo{year}{2007}\natexlab{}.
\newblock \showarticletitle{Computers in talk-based mental health interventions}.
\newblock \bibinfo{journal}{\emph{Interacting with Computers}} \bibinfo{volume}{19}, \bibinfo{number}{4} (\bibinfo{date}{July} \bibinfo{year}{2007}), \bibinfo{pages}{545--562}.
\newblock
\showISSN{09535438}
\href{https://doi.org/10.1016/j.intcom.2007.02.001}{doi:\nolinkurl{10.1016/j.intcom.2007.02.001}}


\bibitem[{Department of Veterans Affairs} and {Department of Defense}(2023)]%
        {VA2023CPG}
\bibfield{author}{\bibinfo{person}{{Department of Veterans Affairs}} {and} \bibinfo{person}{{Department of Defense}}.} \bibinfo{year}{2023}\natexlab{}.
\newblock \bibinfo{title}{VA/DoD Clinical Practice Guideline for Management of Posttraumatic Stress Disorder and Acute Stress Disorder}.
\newblock
\urldef\tempurl%
\url{https://www.healthquality.va.gov/HEALTHQUALITY/guidelines/MH/ptsd/VA-DoD-CPG-PTSD-Full-CPG-Edited-111624-V5-81825.pdf}
\showURL{%
\tempurl}


\bibitem[Ding et~al\mbox{.}(2023)]%
        {ding2023infrastructural}
\bibfield{author}{\bibinfo{person}{Xianghua Ding}, \bibinfo{person}{Linda Tran}, \bibinfo{person}{Yanling Liu}, \bibinfo{person}{Conor O'Neill}, {and} \bibinfo{person}{Stephen Lindsay}.} \bibinfo{year}{2023}\natexlab{}.
\newblock \showarticletitle{Infrastructural Work Behind The Scene: A Study of Formalized Peer-support Practices for Mental Health}. In \bibinfo{booktitle}{\emph{Proceedings of the 2023 CHI Conference on Human Factors in Computing Systems}}. \bibinfo{pages}{1--14}.
\newblock


\bibitem[Evans et~al\mbox{.}(2020)]%
        {evans_understanding_2020}
\bibfield{author}{\bibinfo{person}{Hayley Evans}, \bibinfo{person}{Udaya Lakshmi}, \bibinfo{person}{Hue Watson}, \bibinfo{person}{Azra Ismail}, \bibinfo{person}{Andrew~M. Sherrill}, \bibinfo{person}{Neha Kumar}, {and} \bibinfo{person}{Rosa~I. Arriaga}.} \bibinfo{year}{2020}\natexlab{}.
\newblock \showarticletitle{Understanding the {Care} {Ecologies} of {Veterans} with {PTSD}}. In \bibinfo{booktitle}{\emph{Proceedings of the 2020 {CHI} {Conference} on {Human} {Factors} in {Computing} {Systems}}}. \bibinfo{publisher}{ACM}, \bibinfo{pages}{1--15}.
\newblock
\showISBNx{978-1-4503-6708-0}
\href{https://doi.org/10.1145/3313831.3376170}{doi:\nolinkurl{10.1145/3313831.3376170}}


\bibitem[Evans et~al\mbox{.}(2022)]%
        {evans_perspectives_2022}
\bibfield{author}{\bibinfo{person}{Hayley~Irene Evans}, \bibinfo{person}{Catherine~R Deeter}, \bibinfo{person}{Jiawei Zhou}, \bibinfo{person}{Kimberly Do}, \bibinfo{person}{Andrew~M Sherrill}, {and} \bibinfo{person}{Rosa~I. Arriaga}.} \bibinfo{year}{2022}\natexlab{}.
\newblock \showarticletitle{Perspectives on {Integrating} {Trusted} {Other} {Feedback} in {Therapy} for {Veterans} with {PTSD}}. In \bibinfo{booktitle}{\emph{{CHI} {Conference} on {Human} {Factors} in {Computing} {Systems}}}. \bibinfo{publisher}{ACM}, \bibinfo{pages}{1--16}.
\newblock
\showISBNx{978-1-4503-9157-3}
\href{https://doi.org/10.1145/3491102.3517513}{doi:\nolinkurl{10.1145/3491102.3517513}}


\bibitem[Evans et~al\mbox{.}(2024)]%
        {evansUsingSensorCapturedPatientGenerated2024}
\bibfield{author}{\bibinfo{person}{Hayley~I Evans}, \bibinfo{person}{Myeonghan Ryu}, \bibinfo{person}{Theresa Hsieh}, \bibinfo{person}{Jiawei Zhou}, \bibinfo{person}{Kefan Xu}, \bibinfo{person}{Kenneth~W Akers}, \bibinfo{person}{Andrew~M Sherrill}, {and} \bibinfo{person}{Rosa~I Arriaga}.} \bibinfo{year}{2024}\natexlab{}.
\newblock \showarticletitle{Using {Sensor}-{Captured} {Patient}-{Generated} {Data} to {Support} {Clinical} {Decision}-making in {PTSD} {Therapy}}.
\newblock \bibinfo{journal}{\emph{Proceedings of the ACM on Human-Computer Interaction}}  \bibinfo{volume}{8} (\bibinfo{year}{2024}).
\newblock


\bibitem[Fairburn and Cooper(2011)]%
        {fairburn2011therapist}
\bibfield{author}{\bibinfo{person}{Christopher~G Fairburn} {and} \bibinfo{person}{Zafra Cooper}.} \bibinfo{year}{2011}\natexlab{}.
\newblock \showarticletitle{Therapist competence, therapy quality, and therapist training}.
\newblock \bibinfo{journal}{\emph{Behaviour research and therapy}} \bibinfo{volume}{49}, \bibinfo{number}{6-7} (\bibinfo{year}{2011}).
\newblock


\bibitem[Feinstein(2021)]%
        {feinstein2021descriptions}
\bibfield{author}{\bibinfo{person}{Robert~E Feinstein}.} \bibinfo{year}{2021}\natexlab{}.
\newblock \showarticletitle{Descriptions and reflections on the cognitive apprenticeship model of psychotherapy training \& supervision}.
\newblock \bibinfo{journal}{\emph{Journal of contemporary psychotherapy}} \bibinfo{volume}{51}, \bibinfo{number}{2} (\bibinfo{year}{2021}), \bibinfo{pages}{155--164}.
\newblock


\bibitem[Foa et~al\mbox{.}(2019)]%
        {foa_prolonged_2019}
\bibfield{author}{\bibinfo{person}{Edna~B Foa}, \bibinfo{person}{Elizabeth~Ann Hembree}, \bibinfo{person}{Barbara~Olasov Rothbaum}, {and} \bibinfo{person}{Sheila~AM Rauch}.} \bibinfo{year}{2019}\natexlab{}.
\newblock \bibinfo{booktitle}{\emph{Prolonged exposure therapy for {PTSD}: emotional processing of traumatic experiences: therapist guide} (\bibinfo{edition}{second edition} ed.)}.
\newblock \bibinfo{publisher}{Oxford University Press}, \bibinfo{address}{New York}.
\newblock
\showISBNx{978-0-19-092693-9}
\newblock
\shownote{OCLC: on1111296019}.


\bibitem[Foong et~al\mbox{.}(2018)]%
        {foong_harvesting_2018}
\bibfield{author}{\bibinfo{person}{Pin~Sym Foong}, \bibinfo{person}{Shengdong Zhao}, \bibinfo{person}{Felicia Tan}, {and} \bibinfo{person}{Joseph~Jay Williams}.} \bibinfo{year}{2018}\natexlab{}.
\newblock \showarticletitle{Harvesting {Caregiving} {Knowledge}: {Design} {Considerations} for {Integrating} {Volunteer} {Input} in {Dementia} {Care}}. In \bibinfo{booktitle}{\emph{Proceedings of the 2018 {CHI} {Conference} on {Human} {Factors} in {Computing} {Systems}}}. \bibinfo{publisher}{ACM}, \bibinfo{pages}{1--12}.
\newblock
\showISBNx{978-1-4503-5620-6}
\href{https://doi.org/10.1145/3173574.3173653}{doi:\nolinkurl{10.1145/3173574.3173653}}


\bibitem[Foster and Cone(1995)]%
        {foster_validity_1995}
\bibfield{author}{\bibinfo{person}{Sharon~L Foster} {and} \bibinfo{person}{John~D Cone}.} \bibinfo{year}{1995}\natexlab{}.
\newblock \showarticletitle{Validity {Issues} in {Clinical} {Assessment}}.
\newblock  (\bibinfo{year}{1995}).
\newblock


\bibitem[Frieh(2024)]%
        {frieh2024resistance}
\bibfield{author}{\bibinfo{person}{Ellis~C Frieh}.} \bibinfo{year}{2024}\natexlab{}.
\newblock \showarticletitle{Resistance to the biomedicalization of mental illness through peer support: The case of peer specialists and mental health}.
\newblock \bibinfo{journal}{\emph{Social Science \& Medicine}}  \bibinfo{volume}{341} (\bibinfo{year}{2024}).
\newblock


\bibitem[Frost et~al\mbox{.}(2013)]%
        {frost_supporting_2013}
\bibfield{author}{\bibinfo{person}{Mads Frost}, \bibinfo{person}{Afsaneh Doryab}, \bibinfo{person}{Maria Faurholt-Jepsen}, \bibinfo{person}{Lars~Vedel Kessing}, {and} \bibinfo{person}{Jakob~E Bardram}.} \bibinfo{year}{2013}\natexlab{}.
\newblock \showarticletitle{Supporting disease insight through data analysis: refinements of the monarca self-assessment system}. In \bibinfo{booktitle}{\emph{Proceedings of the 2013 {ACM} international joint conference on {Pervasive} and ubiquitous computing}}. \bibinfo{publisher}{ACM}, \bibinfo{pages}{133--142}.
\newblock
\showISBNx{978-1-4503-1770-2}
\href{https://doi.org/10.1145/2493432.2493507}{doi:\nolinkurl{10.1145/2493432.2493507}}


\bibitem[Furniss and Blandford(2010)]%
        {Furniss2010}
\bibfield{author}{\bibinfo{person}{Dominic Furniss} {and} \bibinfo{person}{Ann Blandford}.} \bibinfo{year}{2010}\natexlab{}.
\newblock \showarticletitle{DiCoT modeling: from analysis to design}.
\newblock  (\bibinfo{year}{2010}).
\newblock


\bibitem[Golden et~al\mbox{.}(2024)]%
        {golden2024applying}
\bibfield{author}{\bibinfo{person}{Grace Golden}, \bibinfo{person}{Christina Popescu}, \bibinfo{person}{Sonia Israel}, \bibinfo{person}{Kelly Perlman}, \bibinfo{person}{Caitrin Armstrong}, \bibinfo{person}{Robert Fratila}, \bibinfo{person}{Myriam Tanguay-Sela}, {and} \bibinfo{person}{David Benrimoh}.} \bibinfo{year}{2024}\natexlab{}.
\newblock \showarticletitle{Applying artificial intelligence to clinical decision support in mental health: What have we learned?}
\newblock \bibinfo{journal}{\emph{Health Policy and Technology}} \bibinfo{volume}{13}, \bibinfo{number}{2} (\bibinfo{year}{2024}), \bibinfo{pages}{100844}.
\newblock


\bibitem[Hetherington(2018)]%
        {hetherington2018infrastructure}
\bibfield{author}{\bibinfo{person}{Kregg Hetherington}.} \bibinfo{year}{2018}\natexlab{}.
\newblock \bibinfo{booktitle}{\emph{Infrastructure, environment, and life in the Anthropocene}}.
\newblock \bibinfo{publisher}{Duke University Press}.
\newblock


\bibitem[Hogan et~al\mbox{.}(2025)]%
        {hogan2025veteran}
\bibfield{author}{\bibinfo{person}{Timothy~P Hogan}, \bibinfo{person}{Bella Etingen}, \bibinfo{person}{Mark~S Zocchi}, \bibinfo{person}{Felicia~R Bixler}, \bibinfo{person}{Nicholas McMahon}, \bibinfo{person}{Jamie Patrianakos}, \bibinfo{person}{Stephanie~A Robinson}, \bibinfo{person}{Terry Newton}, \bibinfo{person}{Nilesh Shah}, \bibinfo{person}{Kathleen~L Frisbee}, {et~al\mbox{.}}} \bibinfo{year}{2025}\natexlab{}.
\newblock \showarticletitle{Veteran Preferences and Willingness to Share Patient-Generated Health Data}.
\newblock \bibinfo{journal}{\emph{Journal of general internal medicine}} \bibinfo{volume}{40}, \bibinfo{number}{5} (\bibinfo{year}{2025}), \bibinfo{pages}{1157--1165}.
\newblock


\bibitem[Hollan et~al\mbox{.}(2000)]%
        {Hollan2000}
\bibfield{author}{\bibinfo{person}{James Hollan}, \bibinfo{person}{Edwin Hutchins}, {and} \bibinfo{person}{David Kirsh}.} \bibinfo{year}{2000}\natexlab{}.
\newblock \showarticletitle{Distributed cognition: toward a new foundation for human-computer interaction research}.
\newblock \bibinfo{journal}{\emph{ACM Transactions on Computer-Human Interaction (TOCHI)}} \bibinfo{volume}{7}, \bibinfo{number}{2} (\bibinfo{year}{2000}), \bibinfo{pages}{174--196}.
\newblock


\bibitem[Hong et~al\mbox{.}(2013)]%
        {hong_investigating_2013}
\bibfield{author}{\bibinfo{person}{Hwajung Hong}, \bibinfo{person}{Svetlana Yarosh}, \bibinfo{person}{Jennifer~G. Kim}, \bibinfo{person}{Gregory~D. Abowd}, {and} \bibinfo{person}{Rosa~I. Arriaga}.} \bibinfo{year}{2013}\natexlab{}.
\newblock \showarticletitle{Investigating the use of circles in social networks to support independence of individuals with autism}. In \bibinfo{booktitle}{\emph{Proceedings of the {SIGCHI} {Conference} on {Human} {Factors} in {Computing} {Systems}}}. \bibinfo{publisher}{ACM}, \bibinfo{pages}{3207--3216}.
\newblock
\showISBNx{978-1-4503-1899-0}
\href{https://doi.org/10.1145/2470654.2466439}{doi:\nolinkurl{10.1145/2470654.2466439}}


\bibitem[Howard et~al\mbox{.}(2023)]%
        {howard2023cognitive}
\bibfield{author}{\bibinfo{person}{Maxine Howard}, \bibinfo{person}{Pippa Hembry}, \bibinfo{person}{Charlotte Rhind}, \bibinfo{person}{Amy Siddall}, \bibinfo{person}{Mohammed~Fahim Uddin}, {and} \bibinfo{person}{Rachel Bryant-Waugh}.} \bibinfo{year}{2023}\natexlab{}.
\newblock \showarticletitle{Cognitive behaviour therapy (CBT) as a psychological intervention in the treatment of ARFID for children and young people}.
\newblock \bibinfo{journal}{\emph{The Cognitive Behaviour Therapist}}  \bibinfo{volume}{16} (\bibinfo{year}{2023}).
\newblock


\bibitem[Hundt et~al\mbox{.}(2020)]%
        {hundt2020didn}
\bibfield{author}{\bibinfo{person}{Natalie~E Hundt}, \bibinfo{person}{Anthony~H Ecker}, \bibinfo{person}{Karin Thompson}, \bibinfo{person}{Ashley Helm}, \bibinfo{person}{Tracey~L Smith}, \bibinfo{person}{Melinda~A Stanley}, {and} \bibinfo{person}{Jeffrey~A Cully}.} \bibinfo{year}{2020}\natexlab{}.
\newblock \showarticletitle{“It didn't fit for me:” A qualitative examination of dropout from prolonged exposure and cognitive processing therapy in veterans.}
\newblock \bibinfo{journal}{\emph{Psychological Services}} \bibinfo{volume}{17}, \bibinfo{number}{4} (\bibinfo{year}{2020}), \bibinfo{pages}{414}.
\newblock


\bibitem[Hussain and Weibel(2016)]%
        {hussain2016can}
\bibfield{author}{\bibinfo{person}{Mustafa Hussain} {and} \bibinfo{person}{Nadir Weibel}.} \bibinfo{year}{2016}\natexlab{}.
\newblock \showarticletitle{Can DiCoT improve infection control? A distributed cognition study of information flow in intensive care}. In \bibinfo{booktitle}{\emph{Proceedings of the 2016 CHI Conference Extended Abstracts on Human Factors in Computing Systems}}. \bibinfo{pages}{2126--2133}.
\newblock


\bibitem[Hutchins(1995)]%
        {Hutchins1995}
\bibfield{author}{\bibinfo{person}{Edwin Hutchins}.} \bibinfo{year}{1995}\natexlab{}.
\newblock \showarticletitle{How a cockpit remembers its speeds}.
\newblock \bibinfo{journal}{\emph{Cognitive science}} \bibinfo{volume}{19}, \bibinfo{number}{3} (\bibinfo{year}{1995}), \bibinfo{pages}{265--288}.
\newblock


\bibitem[Ismail and Kumar(2019)]%
        {ismail2019empowerment}
\bibfield{author}{\bibinfo{person}{Azra Ismail} {and} \bibinfo{person}{Neha Kumar}.} \bibinfo{year}{2019}\natexlab{}.
\newblock \showarticletitle{Empowerment on the margins: The online experiences of community health workers}. In \bibinfo{booktitle}{\emph{Proceedings of the 2019 CHI Conference on Human Factors in Computing Systems}}. \bibinfo{pages}{1--15}.
\newblock


\bibitem[J{\"a}{\"a}skel{\"a}inen(2010)]%
        {jaaskelainen2010think}
\bibfield{author}{\bibinfo{person}{Riitta J{\"a}{\"a}skel{\"a}inen}.} \bibinfo{year}{2010}\natexlab{}.
\newblock \showarticletitle{Think-aloud protocol}.
\newblock \bibinfo{journal}{\emph{Handbook of translation studies}}  \bibinfo{volume}{1} (\bibinfo{year}{2010}), \bibinfo{pages}{371--374}.
\newblock


\bibitem[Jacobs et~al\mbox{.}(2021)]%
        {jacobs_designing_2021}
\bibfield{author}{\bibinfo{person}{Maia Jacobs}, \bibinfo{person}{Jeffrey He}, \bibinfo{person}{Melanie~F Pradier}, \bibinfo{person}{Barbara Lam}, \bibinfo{person}{Andrew~C Ahn}, \bibinfo{person}{Thomas~H McCoy}, \bibinfo{person}{Roy~H Perlis}, \bibinfo{person}{Finale Doshi-Velez}, {and} \bibinfo{person}{Krzysztof~Z Gajos}.} \bibinfo{year}{2021}\natexlab{}.
\newblock \showarticletitle{Designing {AI} for {Trust} and {Collaboration} in {Time}-{Constrained} {Medical} {Decisions}: {A} {Sociotechnical} {Lens}}. In \bibinfo{booktitle}{\emph{Proceedings of the 2021 {CHI} {Conference} on {Human} {Factors} in {Computing} {Systems}}}. \bibinfo{pages}{1--14}.
\newblock
\href{https://doi.org/10.1145/3411764.3445385}{doi:\nolinkurl{10.1145/3411764.3445385}}


\bibitem[Jacobs et~al\mbox{.}(2015)]%
        {jacobs_comparing_2015}
\bibfield{author}{\bibinfo{person}{Maia~L. Jacobs}, \bibinfo{person}{James Clawson}, {and} \bibinfo{person}{Elizabeth~D. Mynatt}.} \bibinfo{year}{2015}\natexlab{}.
\newblock \showarticletitle{Comparing {Health} {Information} {Sharing} {Preferences} of {Cancer} {Patients}, {Doctors}, and {Navigators}}. In \bibinfo{booktitle}{\emph{Proceedings of the 18th {ACM} {Conference} on {Computer} {Supported} {Cooperative} {Work} \& {Social} {Computing}}}. \bibinfo{publisher}{ACM}, \bibinfo{pages}{808--818}.
\newblock
\showISBNx{978-1-4503-2922-4}
\href{https://doi.org/10.1145/2675133.2675252}{doi:\nolinkurl{10.1145/2675133.2675252}}


\bibitem[Javaheri et~al\mbox{.}(2025)]%
        {javaheri2025concept}
\bibfield{author}{\bibinfo{person}{Hamraz Javaheri}, \bibinfo{person}{Omid Ghamarnejad}, \bibinfo{person}{Paul Lukowicz}, \bibinfo{person}{Gregor~A Stavrou}, {and} \bibinfo{person}{Jakob Karolus}.} \bibinfo{year}{2025}\natexlab{}.
\newblock \showarticletitle{From Concept to Clinic: Multidisciplinary Design, Development, and Clinical Validation of Augmented Reality-Assisted Open Pancreatic Surgery}. In \bibinfo{booktitle}{\emph{Proceedings of the 2025 CHI Conference on Human Factors in Computing Systems}}. \bibinfo{pages}{1--24}.
\newblock


\bibitem[Jin et~al\mbox{.}(2017)]%
        {jin2017collaborative}
\bibfield{author}{\bibinfo{person}{Weina Jin}, \bibinfo{person}{Diane Gromala}, \bibinfo{person}{Carman Neustaedter}, {and} \bibinfo{person}{Xin Tong}.} \bibinfo{year}{2017}\natexlab{}.
\newblock \showarticletitle{A Collaborative Visualization Tool to Support Doctors' Shared Decision-Making on Antibiotic Prescription}. In \bibinfo{booktitle}{\emph{Companion of the 2017 ACM Conference on Computer Supported Cooperative Work and Social Computing}}. \bibinfo{pages}{211--214}.
\newblock


\bibitem[Jo et~al\mbox{.}(2022)]%
        {jo_designing_2022}
\bibfield{author}{\bibinfo{person}{Eunkyung Jo}, \bibinfo{person}{Myeonghan Ryu}, \bibinfo{person}{Georgia Kenderova}, \bibinfo{person}{Samuel So}, \bibinfo{person}{Bryan Shapiro}, \bibinfo{person}{Alexandra Papoutsaki}, {and} \bibinfo{person}{Daniel~A. Epstein}.} \bibinfo{year}{2022}\natexlab{}.
\newblock \showarticletitle{Designing {Flexible} {Longitudinal} {Regimens}: {Supporting} {Clinician} {Planning} for {Discontinuation} of {Psychiatric} {Drugs}}. In \bibinfo{booktitle}{\emph{{CHI} {Conference} on {Human} {Factors} in {Computing} {Systems}}}. \bibinfo{publisher}{ACM}, \bibinfo{pages}{1--17}.
\newblock
\showISBNx{978-1-4503-9157-3}
\href{https://doi.org/10.1145/3491102.3502206}{doi:\nolinkurl{10.1145/3491102.3502206}}


\bibitem[Juarascio et~al\mbox{.}(2020)]%
        {juarascio_momentary_2020}
\bibfield{author}{\bibinfo{person}{Adrienne~S Juarascio}, \bibinfo{person}{Rebecca~J Crochiere}, \bibinfo{person}{Tinashe~M Tapera}, \bibinfo{person}{Madeline Palermo}, {and} \bibinfo{person}{Fengqing Zhang}.} \bibinfo{year}{2020}\natexlab{}.
\newblock \showarticletitle{Momentary changes in heart rate variability can detect risk for emotional eating episodes}.
\newblock \bibinfo{journal}{\emph{Appetite}}  \bibinfo{volume}{152} (\bibinfo{date}{Sept.} \bibinfo{year}{2020}), \bibinfo{pages}{104698}.
\newblock
\showISSN{01956663}
\href{https://doi.org/10.1016/j.appet.2020.104698}{doi:\nolinkurl{10.1016/j.appet.2020.104698}}


\bibitem[Karter(2025)]%
        {karter2025ecological}
\bibfield{author}{\bibinfo{person}{Justin~M Karter}.} \bibinfo{year}{2025}\natexlab{}.
\newblock \showarticletitle{An ecological model for conceptual competence in psychiatric diagnosis}.
\newblock \bibinfo{journal}{\emph{Journal of Humanistic Psychology}} \bibinfo{volume}{65}, \bibinfo{number}{4} (\bibinfo{year}{2025}), \bibinfo{pages}{741--766}.
\newblock


\bibitem[Kawanishi et~al\mbox{.}(2015)]%
        {kawanishi_lifelog-based_2015}
\bibfield{author}{\bibinfo{person}{Nao Kawanishi}, \bibinfo{person}{Morihiko Tamai}, \bibinfo{person}{Akio Hasegawa}, \bibinfo{person}{Yoshio Takeuchi}, \bibinfo{person}{Aran Tajika}, \bibinfo{person}{Yusuke Ogawa}, {and} \bibinfo{person}{Toshiaki Furukawa}.} \bibinfo{year}{2015}\natexlab{}.
\newblock \showarticletitle{Lifelog-based estimation of activity diary for cognitive behavioral therapy}. In \bibinfo{booktitle}{\emph{Proceedings of the 2015 {ACM} {International} {Joint} {Conference} on {Pervasive} and {Ubiquitous} {Computing} and {Proceedings} of the 2015 {ACM} {International} {Symposium} on {Wearable} {Computers}}}. \bibinfo{publisher}{ACM Press}, \bibinfo{pages}{1251--1256}.
\newblock
\showISBNx{978-1-4503-3575-1}
\href{https://doi.org/10.1145/2800835.2807939}{doi:\nolinkurl{10.1145/2800835.2807939}}


\bibitem[Kazantzis et~al\mbox{.}(2010)]%
        {kazantzis2010meta}
\bibfield{author}{\bibinfo{person}{Nikolaos Kazantzis}, \bibinfo{person}{Craig Whittington}, {and} \bibinfo{person}{Frank Dattilio}.} \bibinfo{year}{2010}\natexlab{}.
\newblock \showarticletitle{Meta-analysis of homework effects in cognitive and behavioral therapy: A replication and extension.}
\newblock \bibinfo{journal}{\emph{Clinical Psychology: Science and Practice}} \bibinfo{volume}{17}, \bibinfo{number}{2} (\bibinfo{year}{2010}), \bibinfo{pages}{144}.
\newblock


\bibitem[Kazdin(2017)]%
        {kazdinAddressingTreatmentGap2017a}
\bibfield{author}{\bibinfo{person}{Alan~E Kazdin}.} \bibinfo{year}{2017}\natexlab{}.
\newblock \showarticletitle{Addressing the treatment gap: {A} key challenge for extending evidence-based psychosocial interventions}.
\newblock \bibinfo{journal}{\emph{Behaviour Research and Therapy}}  \bibinfo{volume}{88} (\bibinfo{date}{Jan.} \bibinfo{year}{2017}), \bibinfo{pages}{7--18}.
\newblock
\showISSN{00057967}
\href{https://doi.org/10.1016/j.brat.2016.06.004}{doi:\nolinkurl{10.1016/j.brat.2016.06.004}}


\bibitem[Kehle-Forbes et~al\mbox{.}(2022)]%
        {kehle2022divergent}
\bibfield{author}{\bibinfo{person}{Shannon~M Kehle-Forbes}, \bibinfo{person}{Princess~E Ackland}, \bibinfo{person}{Michele~R Spoont}, \bibinfo{person}{Laura~A Meis}, \bibinfo{person}{Robert~J Orazem}, \bibinfo{person}{Alexandra Lyon}, \bibinfo{person}{Helen~R Valenstein-Mah}, \bibinfo{person}{Paula~P Schnurr}, \bibinfo{person}{Susan~L Zickmund}, \bibinfo{person}{Edna~B Foa}, {et~al\mbox{.}}} \bibinfo{year}{2022}\natexlab{}.
\newblock \showarticletitle{Divergent experiences of US veterans who did and did not complete trauma-focused therapies for PTSD: A national qualitative study of treatment dropout}.
\newblock \bibinfo{journal}{\emph{Behaviour Research and Therapy}}  \bibinfo{volume}{154} (\bibinfo{year}{2022}), \bibinfo{pages}{104123}.
\newblock


\bibitem[Kelley et~al\mbox{.}(2017)]%
        {kelley_self-tracking_2017}
\bibfield{author}{\bibinfo{person}{Christina Kelley}, \bibinfo{person}{Bongshin Lee}, {and} \bibinfo{person}{Lauren Wilcox}.} \bibinfo{year}{2017}\natexlab{}.
\newblock \showarticletitle{Self-tracking for {Mental} {Wellness}: {Understanding} {Expert} {Perspectives} and {Student} {Experiences}}. In \bibinfo{booktitle}{\emph{Proceedings of the 2017 {CHI} {Conference} on {Human} {Factors} in {Computing} {Systems}}}. \bibinfo{pages}{629--641}.
\newblock
\showISBNx{978-1-4503-4655-9}
\href{https://doi.org/10.1145/3025453.3025750}{doi:\nolinkurl{10.1145/3025453.3025750}}


\bibitem[Khairat et~al\mbox{.}(2018)]%
        {khairat_reasons_2018}
\bibfield{author}{\bibinfo{person}{Saif Khairat}, \bibinfo{person}{David Marc}, \bibinfo{person}{William Crosby}, {and} \bibinfo{person}{Ali Al~Sanousi}.} \bibinfo{year}{2018}\natexlab{}.
\newblock \showarticletitle{Reasons {For} {Physicians} {Not} {Adopting} {Clinical} {Decision} {Support} {Systems}: {Critical} {Analysis}}.
\newblock \bibinfo{journal}{\emph{JMIR Medical Informatics}} \bibinfo{volume}{6}, \bibinfo{number}{2} (\bibinfo{date}{April} \bibinfo{year}{2018}).
\newblock
\showISSN{2291-9694}
\href{https://doi.org/10.2196/medinform.8912}{doi:\nolinkurl{10.2196/medinform.8912}}


\bibitem[Khatiwada et~al\mbox{.}(2024)]%
        {khatiwada2024patient}
\bibfield{author}{\bibinfo{person}{Pankaj Khatiwada}, \bibinfo{person}{Bian Yang}, \bibinfo{person}{Jia-Chun Lin}, {and} \bibinfo{person}{Bernd Blobel}.} \bibinfo{year}{2024}\natexlab{}.
\newblock \showarticletitle{Patient-generated health data (PGHD): understanding, requirements, challenges, and existing techniques for data security and privacy}.
\newblock \bibinfo{journal}{\emph{Journal of personalized medicine}} \bibinfo{volume}{14}, \bibinfo{number}{3} (\bibinfo{year}{2024}), \bibinfo{pages}{282}.
\newblock


\bibitem[Kl{\"u}ber et~al\mbox{.}(2020)]%
        {kluber2020experience}
\bibfield{author}{\bibinfo{person}{Sara Kl{\"u}ber}, \bibinfo{person}{Franzisca Maas}, \bibinfo{person}{David Schraudt}, \bibinfo{person}{Gina Hermann}, \bibinfo{person}{Oliver Happel}, {and} \bibinfo{person}{Tobias Grundgeiger}.} \bibinfo{year}{2020}\natexlab{}.
\newblock \showarticletitle{Experience matters: design and evaluation of an anesthesia support tool guided by user experience theory}. In \bibinfo{booktitle}{\emph{Proceedings of the 2020 ACM Designing Interactive Systems Conference}}. \bibinfo{pages}{1523--1535}.
\newblock


\bibitem[Kouri et~al\mbox{.}(2022)]%
        {kouri_providers_2022}
\bibfield{author}{\bibinfo{person}{Andrew Kouri}, \bibinfo{person}{Janet Yamada}, \bibinfo{person}{Jeffrey Lam Shin~Cheung}, \bibinfo{person}{Stijn Van~de Velde}, {and} \bibinfo{person}{Samir Gupta}.} \bibinfo{year}{2022}\natexlab{}.
\newblock \showarticletitle{Do providers use computerized clinical decision support systems? {A} systematic review and meta-regression of clinical decision support uptake}.
\newblock \bibinfo{journal}{\emph{Implementation Science}} \bibinfo{volume}{17}, \bibinfo{number}{1} (\bibinfo{date}{Dec.} \bibinfo{year}{2022}), \bibinfo{pages}{21}.
\newblock
\showISSN{1748-5908}
\href{https://doi.org/10.1186/s13012-022-01199-3}{doi:\nolinkurl{10.1186/s13012-022-01199-3}}


\bibitem[Kuester et~al\mbox{.}(2016)]%
        {kuester2016internet}
\bibfield{author}{\bibinfo{person}{Annika Kuester}, \bibinfo{person}{Helen Niemeyer}, {and} \bibinfo{person}{Christine Knaevelsrud}.} \bibinfo{year}{2016}\natexlab{}.
\newblock \showarticletitle{Internet-based interventions for posttraumatic stress: A meta-analysis of randomized controlled trials}.
\newblock \bibinfo{journal}{\emph{Clinical psychology review}}  \bibinfo{volume}{43} (\bibinfo{year}{2016}), \bibinfo{pages}{1--16}.
\newblock


\bibitem[Kuhn et~al\mbox{.}(2017)]%
        {kuhn_randomized_2017}
\bibfield{author}{\bibinfo{person}{Eric Kuhn}, \bibinfo{person}{Nitya Kanuri}, \bibinfo{person}{Julia~E Hoffman}, \bibinfo{person}{Donn~W Garvert}, \bibinfo{person}{Josef~I Ruzek}, {and} \bibinfo{person}{C~Barr Taylor}.} \bibinfo{year}{2017}\natexlab{}.
\newblock \showarticletitle{A randomized controlled trial of a smartphone app for posttraumatic stress disorder symptoms.}
\newblock \bibinfo{journal}{\emph{Journal of Consulting and Clinical Psychology}} \bibinfo{volume}{85}, \bibinfo{number}{3} (\bibinfo{date}{March} \bibinfo{year}{2017}), \bibinfo{pages}{267--273}.
\newblock
\showISSN{1939-2117, 0022-006X}
\href{https://doi.org/10.1037/ccp0000163}{doi:\nolinkurl{10.1037/ccp0000163}}


\bibitem[Lave and Wenger(1991)]%
        {lave1991situated}
\bibfield{author}{\bibinfo{person}{Jean Lave} {and} \bibinfo{person}{Etienne Wenger}.} \bibinfo{year}{1991}\natexlab{}.
\newblock \bibinfo{booktitle}{\emph{Situated learning: Legitimate peripheral participation}}.
\newblock \bibinfo{publisher}{Cambridge university press}.
\newblock


\bibitem[Lee et~al\mbox{.}(2022)]%
        {lee_towards_2022}
\bibfield{author}{\bibinfo{person}{Min~Hun Lee}, \bibinfo{person}{Daniel~P Siewiorek}, \bibinfo{person}{Asim Smailagic}, \bibinfo{person}{Alexandre Bernardino}, {and} \bibinfo{person}{Sergi Bermúdez I~Badia}.} \bibinfo{year}{2022}\natexlab{}.
\newblock \showarticletitle{Towards {Efficient} {Annotations} for a {Human}-{AI} {Collaborative}, {Clinical} {Decision} {Support} {System}: {A} {Case} {Study} on {Physical} {Stroke} {Rehabilitation} {Assessment}}. In \bibinfo{booktitle}{\emph{27th {International} {Conference} on {Intelligent} {User} {Interfaces}}}. \bibinfo{publisher}{ACM}, \bibinfo{address}{Helsinki Finland}, \bibinfo{pages}{4--14}.
\newblock
\showISBNx{978-1-4503-9144-3}
\href{https://doi.org/10.1145/3490099.3511112}{doi:\nolinkurl{10.1145/3490099.3511112}}


\bibitem[Lee et~al\mbox{.}(2021)]%
        {lee_human-ai_2021}
\bibfield{author}{\bibinfo{person}{Min~Hun Lee}, \bibinfo{person}{Daniel~PP Siewiorek}, \bibinfo{person}{Asim Smailagic}, \bibinfo{person}{Alexandre Bernardino}, {and} \bibinfo{person}{Sergi~Bermúdez Bermúdez~i Badia}.} \bibinfo{year}{2021}\natexlab{}.
\newblock \showarticletitle{A {Human}-{AI} {Collaborative} {Approach} for {Clinical} {Decision} {Making} on {Rehabilitation} {Assessment}}. In \bibinfo{booktitle}{\emph{Proceedings of the 2021 {CHI} {Conference} on {Human} {Factors} in {Computing} {Systems}}}. \bibinfo{publisher}{ACM}, \bibinfo{pages}{1--14}.
\newblock
\showISBNx{978-1-4503-8096-6}
\href{https://doi.org/10.1145/3411764.3445472}{doi:\nolinkurl{10.1145/3411764.3445472}}


\bibitem[Leist et~al\mbox{.}(2025)]%
        {leist2025towards}
\bibfield{author}{\bibinfo{person}{Robert~Andreas Leist}, \bibinfo{person}{Hans-J{\"u}rgen Profitlich}, \bibinfo{person}{Tim Hunsicker}, {and} \bibinfo{person}{Daniel Sonntag}.} \bibinfo{year}{2025}\natexlab{}.
\newblock \showarticletitle{Towards Trustable Intelligent Clinical Decision Support Systems: A User Study with Ophthalmologists}. In \bibinfo{booktitle}{\emph{Proceedings of the 30th International Conference on Intelligent User Interfaces}}. \bibinfo{pages}{1470--1484}.
\newblock


\bibitem[Li et~al\mbox{.}(2010)]%
        {li2010stage}
\bibfield{author}{\bibinfo{person}{Ian Li}, \bibinfo{person}{Anind Dey}, {and} \bibinfo{person}{Jodi Forlizzi}.} \bibinfo{year}{2010}\natexlab{}.
\newblock \showarticletitle{A stage-based model of personal informatics systems}. In \bibinfo{booktitle}{\emph{Proceedings of the SIGCHI conference on human factors in computing systems}}. \bibinfo{pages}{557--566}.
\newblock


\bibitem[Luborsky and Barber(1993)]%
        {luborsky1993benefits}
\bibfield{author}{\bibinfo{person}{Lester Luborsky} {and} \bibinfo{person}{Jacques~P Barber}.} \bibinfo{year}{1993}\natexlab{}.
\newblock \showarticletitle{Benefits of adherence to psychotherapy manuals, and where to get them.}
\newblock  (\bibinfo{year}{1993}).
\newblock


\bibitem[Luborsky and DeRubeis(1984)]%
        {luborsky1984use}
\bibfield{author}{\bibinfo{person}{Lester Luborsky} {and} \bibinfo{person}{Robert~J DeRubeis}.} \bibinfo{year}{1984}\natexlab{}.
\newblock \showarticletitle{The use of psychotherapy treatment manuals: A small revolution in psychotherapy research style}.
\newblock \bibinfo{journal}{\emph{Clinical Psychology Review}} \bibinfo{volume}{4}, \bibinfo{number}{1} (\bibinfo{year}{1984}), \bibinfo{pages}{5--14}.
\newblock


\bibitem[Luxon(2015)]%
        {luxon2015infrastructure}
\bibfield{author}{\bibinfo{person}{Linda Luxon}.} \bibinfo{year}{2015}\natexlab{}.
\newblock \showarticletitle{Infrastructure--the key to healthcare improvement}.
\newblock \bibinfo{journal}{\emph{Future Healthcare Journal}} \bibinfo{volume}{2}, \bibinfo{number}{1} (\bibinfo{year}{2015}), \bibinfo{pages}{4--7}.
\newblock


\bibitem[Maguen et~al\mbox{.}(2019)]%
        {maguen2019factors}
\bibfield{author}{\bibinfo{person}{Shira Maguen}, \bibinfo{person}{Yongmei Li}, \bibinfo{person}{Erin Madden}, \bibinfo{person}{Karen~H Seal}, \bibinfo{person}{Thomas~C Neylan}, \bibinfo{person}{Olga~V Patterson}, \bibinfo{person}{Scott~L DuVall}, \bibinfo{person}{Callan Lujan}, {and} \bibinfo{person}{Brian Shiner}.} \bibinfo{year}{2019}\natexlab{}.
\newblock \showarticletitle{Factors associated with completing evidence-based psychotherapy for PTSD among veterans in a national healthcare system}.
\newblock \bibinfo{journal}{\emph{Psychiatry Research}}  \bibinfo{volume}{274} (\bibinfo{year}{2019}), \bibinfo{pages}{112--128}.
\newblock


\bibitem[Maieritsch et~al\mbox{.}(2016)]%
        {maieritsch2016randomized}
\bibfield{author}{\bibinfo{person}{Kelly~P Maieritsch}, \bibinfo{person}{Tracey~L Smith}, \bibinfo{person}{Jonathan~D Hessinger}, \bibinfo{person}{Eileen~P Ahearn}, \bibinfo{person}{Jens~C Eickhoff}, {and} \bibinfo{person}{Qianqian Zhao}.} \bibinfo{year}{2016}\natexlab{}.
\newblock \showarticletitle{Randomized controlled equivalence trial comparing videoconference and in person delivery of cognitive processing therapy for PTSD}.
\newblock \bibinfo{journal}{\emph{Journal of Telemedicine and Telecare}} \bibinfo{volume}{22}, \bibinfo{number}{4} (\bibinfo{year}{2016}), \bibinfo{pages}{238--243}.
\newblock


\bibitem[Marathe and Toyama(2021)]%
        {marathe2021situated}
\bibfield{author}{\bibinfo{person}{Megh Marathe} {and} \bibinfo{person}{Kentaro Toyama}.} \bibinfo{year}{2021}\natexlab{}.
\newblock \showarticletitle{The situated, relational, and evolving nature of epilepsy diagnosis}.
\newblock \bibinfo{journal}{\emph{Proceedings of the ACM on Human-Computer Interaction}} \bibinfo{volume}{4}, \bibinfo{number}{CSCW3} (\bibinfo{year}{2021}), \bibinfo{pages}{1--18}.
\newblock


\bibitem[Martino et~al\mbox{.}(2016)]%
        {martino2016effectiveness}
\bibfield{author}{\bibinfo{person}{Steve Martino}, \bibinfo{person}{Manuel Paris~Jr}, \bibinfo{person}{Luis A{\~n}ez}, \bibinfo{person}{Charla Nich}, \bibinfo{person}{Monica Canning-Ball}, \bibinfo{person}{Karen Hunkele}, \bibinfo{person}{Todd~A Olmstead}, {and} \bibinfo{person}{Kathleen~M Carroll}.} \bibinfo{year}{2016}\natexlab{}.
\newblock \showarticletitle{The effectiveness and cost of clinical supervision for motivational interviewing: a randomized controlled trial}.
\newblock \bibinfo{journal}{\emph{Journal of Substance Abuse Treatment}}  \bibinfo{volume}{68} (\bibinfo{year}{2016}), \bibinfo{pages}{11--23}.
\newblock


\bibitem[Matthews and Doherty(2011)]%
        {matthews_mood_2011}
\bibfield{author}{\bibinfo{person}{Mark Matthews} {and} \bibinfo{person}{Gavin Doherty}.} \bibinfo{year}{2011}\natexlab{}.
\newblock \showarticletitle{In the mood: engaging teenagers in psychotherapy using mobile phones}. In \bibinfo{booktitle}{\emph{Proceedings of the {SIGCHI} {Conference} on {Human} {Factors} in {Computing} {Systems}}}. \bibinfo{publisher}{ACM}, \bibinfo{pages}{2947--2956}.
\newblock
\showISBNx{978-1-4503-0228-9}
\href{https://doi.org/10.1145/1978942.1979379}{doi:\nolinkurl{10.1145/1978942.1979379}}


\bibitem[Mausbach et~al\mbox{.}(2010)]%
        {mausbach2010relationship}
\bibfield{author}{\bibinfo{person}{Brent~T Mausbach}, \bibinfo{person}{Raeanne Moore}, \bibinfo{person}{Scott Roesch}, \bibinfo{person}{Veronica Cardenas}, {and} \bibinfo{person}{Thomas~L Patterson}.} \bibinfo{year}{2010}\natexlab{}.
\newblock \showarticletitle{The relationship between homework compliance and therapy outcomes: An updated meta-analysis}.
\newblock \bibinfo{journal}{\emph{Cognitive therapy and research}} \bibinfo{volume}{34}, \bibinfo{number}{5} (\bibinfo{year}{2010}), \bibinfo{pages}{429--438}.
\newblock


\bibitem[Mentis et~al\mbox{.}(2017)]%
        {mentis_crafting_2017}
\bibfield{author}{\bibinfo{person}{Helena~M. Mentis}, \bibinfo{person}{Anita Komlodi}, \bibinfo{person}{Katrina Schrader}, \bibinfo{person}{Michael Phipps}, \bibinfo{person}{Ann Gruber-Baldini}, \bibinfo{person}{Karen Yarbrough}, {and} \bibinfo{person}{Lisa Shulman}.} \bibinfo{year}{2017}\natexlab{}.
\newblock \showarticletitle{Crafting a {View} of {Self}-{Tracking} {Data} in the {Clinical} {Visit}}. In \bibinfo{booktitle}{\emph{Proceedings of the 2017 {CHI} {Conference} on {Human} {Factors} in {Computing} {Systems}}}.
\newblock
\showISBNx{978-1-4503-4655-9}
\href{https://doi.org/10.1145/3025453.3025589}{doi:\nolinkurl{10.1145/3025453.3025589}}


\bibitem[Middleton et~al\mbox{.}(2016)]%
        {middleton_clinical_2016}
\bibfield{author}{\bibinfo{person}{B Middleton}, \bibinfo{person}{DF Sittig}, {and} \bibinfo{person}{A Wright}.} \bibinfo{year}{2016}\natexlab{}.
\newblock \showarticletitle{Clinical {Decision} {Support}: a 25 {Year} {Retrospective} and a 25 {Year} {Vision}}.
\newblock \bibinfo{journal}{\emph{Yearbook of Medical Informatics}}  \bibinfo{volume}{25} (\bibinfo{date}{Aug.} \bibinfo{year}{2016}).
\newblock
\showISSN{0943-4747, 2364-0502}
\href{https://doi.org/10.15265/IYS-2016-s034}{doi:\nolinkurl{10.15265/IYS-2016-s034}}


\bibitem[Mishra et~al\mbox{.}(2018)]%
        {mishra2018supporting}
\bibfield{author}{\bibinfo{person}{Sonali~R Mishra}, \bibinfo{person}{Andrew~D Miller}, \bibinfo{person}{Shefali Haldar}, \bibinfo{person}{Maher Khelifi}, \bibinfo{person}{Jordan Eschler}, \bibinfo{person}{Rashmi~G Elera}, \bibinfo{person}{Ari~H Pollack}, {and} \bibinfo{person}{Wanda Pratt}.} \bibinfo{year}{2018}\natexlab{}.
\newblock \showarticletitle{Supporting collaborative health tracking in the hospital: patients' perspectives}. In \bibinfo{booktitle}{\emph{Proceedings of the 2018 CHI Conference on Human Factors in Computing Systems}}. \bibinfo{pages}{1--14}.
\newblock


\bibitem[Morand et~al\mbox{.}(2022)]%
        {Morand2022}
\bibfield{author}{\bibinfo{person}{Oph{\'e}lie Morand}, \bibinfo{person}{St{\'e}phane Safin}, \bibinfo{person}{Caroline Rizza}, \bibinfo{person}{Robert Larribau}, {and} \bibinfo{person}{Romain Pages}.} \bibinfo{year}{2022}\natexlab{}.
\newblock \showarticletitle{Analyzing the challenges of an assistive application’integration in a complex emergency interaction using a distributed cognition perspective}. In \bibinfo{booktitle}{\emph{Proceedings of the 33rd European Conference on Cognitive Ergonomics}}. \bibinfo{pages}{1--9}.
\newblock


\bibitem[Murnane et~al\mbox{.}(2018)]%
        {murnane_personal_2018}
\bibfield{author}{\bibinfo{person}{Elizabeth~L. Murnane}, \bibinfo{person}{Tara~G. Walker}, \bibinfo{person}{Beck Tench}, \bibinfo{person}{Stephen Voida}, {and} \bibinfo{person}{Jaime Snyder}.} \bibinfo{year}{2018}\natexlab{}.
\newblock \showarticletitle{Personal {Informatics} in {Interpersonal} {Contexts}: {Towards} the {Design} of {Technology} that {Supports} the {Social} {Ecologies} of {Long}-{Term} {Mental} {Health} {Management}}.
\newblock \bibinfo{journal}{\emph{Proceedings of the ACM on Human-Computer Interaction}} \bibinfo{volume}{2}, \bibinfo{number}{CSCW} (\bibinfo{date}{Nov.} \bibinfo{year}{2018}), \bibinfo{pages}{1--27}.
\newblock
\showISSN{2573-0142}
\href{https://doi.org/10.1145/3274396}{doi:\nolinkurl{10.1145/3274396}}
\newblock
\shownote{Refer to for deployment paper}.


\bibitem[Neff and Nafus(2016)]%
        {neff2016self}
\bibfield{author}{\bibinfo{person}{Gina Neff} {and} \bibinfo{person}{Dawn Nafus}.} \bibinfo{year}{2016}\natexlab{}.
\newblock \bibinfo{booktitle}{\emph{Self-tracking}}.
\newblock \bibinfo{publisher}{Mit Press}.
\newblock


\bibitem[Ng et~al\mbox{.}(2019)]%
        {ng2019provider}
\bibfield{author}{\bibinfo{person}{Ada Ng}, \bibinfo{person}{Rachel Kornfield}, \bibinfo{person}{Stephen~M Schueller}, \bibinfo{person}{Alyson~K Zalta}, \bibinfo{person}{Michael Brennan}, {and} \bibinfo{person}{Madhu Reddy}.} \bibinfo{year}{2019}\natexlab{}.
\newblock \showarticletitle{Provider perspectives on integrating sensor-captured patient-generated data in mental health care}.
\newblock \bibinfo{journal}{\emph{Proceedings of the ACM on human-computer interaction}} \bibinfo{volume}{3}, \bibinfo{number}{CSCW} (\bibinfo{year}{2019}), \bibinfo{pages}{1--25}.
\newblock


\bibitem[Nie et~al\mbox{.}(2025)]%
        {nie2024llm}
\bibfield{author}{\bibinfo{person}{Jingping Nie}, \bibinfo{person}{Hanya Shao}, \bibinfo{person}{Yuang Fan}, \bibinfo{person}{Qijia Shao}, \bibinfo{person}{Haoxuan You}, \bibinfo{person}{Matthias Preindl}, {and} \bibinfo{person}{Xiaofan Jiang}.} \bibinfo{year}{2025}\natexlab{}.
\newblock \showarticletitle{LLM-based conversational AI therapist for daily functioning screening and psychotherapeutic intervention via everyday smart devices}.
\newblock \bibinfo{journal}{\emph{ACM Transactions on Computing for Healthcare}} (\bibinfo{year}{2025}).
\newblock


\bibitem[Nova et~al\mbox{.}(2024)]%
        {nova2024unveiling}
\bibfield{author}{\bibinfo{person}{Fayika~Farhat Nova}, \bibinfo{person}{Rachel Pfafman}, \bibinfo{person}{Caralyn Logan~Delaney}, {and} \bibinfo{person}{Jessica Pater}.} \bibinfo{year}{2024}\natexlab{}.
\newblock \showarticletitle{Unveiling the ``Toxic'' World of \#Meanspo: Understanding Users' Emerging Online Eating Disorder Practices in X/Twitter}.
\newblock \bibinfo{journal}{\emph{Proceedings of the ACM on Human-Computer Interaction}} \bibinfo{volume}{8}, \bibinfo{number}{CSCW2} (\bibinfo{year}{2024}), \bibinfo{pages}{1--27}.
\newblock


\bibitem[O'Connor et~al\mbox{.}(2018)]%
        {oconnorIncreasingAvailabilityPsychological2018}
\bibfield{author}{\bibinfo{person}{Marianne O'Connor}, \bibinfo{person}{Katy~E Morgan}, \bibinfo{person}{Suzanne Bailey-Straebler}, \bibinfo{person}{Christopher~G Fairburn}, {and} \bibinfo{person}{Zafra Cooper}.} \bibinfo{year}{2018}\natexlab{}.
\newblock \showarticletitle{Increasing the {Availability} of {Psychological} {Treatments}: {A} {Multinational} {Study} of a {Scalable} {Method} for {Training} {Therapists}}.
\newblock \bibinfo{journal}{\emph{Journal of Medical Internet Research}} \bibinfo{volume}{20}, \bibinfo{number}{6} (\bibinfo{date}{June} \bibinfo{year}{2018}).
\newblock
\showISSN{1438-8871}
\href{https://doi.org/10.2196/10386}{doi:\nolinkurl{10.2196/10386}}


\bibitem[of~Veterans~Affairs(2025a)]%
        {VaPtsdPrevalence}
\bibfield{author}{\bibinfo{person}{U.S.~Department of Veterans~Affairs}.} \bibinfo{year}{2025}\natexlab{a}.
\newblock \bibinfo{title}{How Common Is PTSD in Adults?}
\newblock
\urldef\tempurl%
\url{https://www.ptsd.va.gov/understand/common/common_adults.asp}
\showURL{%
Retrieved 2025-08-11 from \tempurl}


\bibitem[of~Veterans~Affairs(2025b)]%
        {VHAStats}
\bibfield{author}{\bibinfo{person}{U.S.~Department of Veterans~Affairs}.} \bibinfo{year}{2025}\natexlab{b}.
\newblock \bibinfo{title}{Veterans Health Administration}.
\newblock
\urldef\tempurl%
\url{https://www.va.gov/health/aboutVHA.asp}
\showURL{%
Retrieved 2025-08-11 from \tempurl}


\bibitem[Oh et~al\mbox{.}(2022)]%
        {oh_patients_2022}
\bibfield{author}{\bibinfo{person}{Chi~Young Oh}, \bibinfo{person}{Yuhan Luo}, \bibinfo{person}{Beth St.~Jean}, {and} \bibinfo{person}{Eun~Kyoung Choe}.} \bibinfo{year}{2022}\natexlab{}.
\newblock \showarticletitle{Patients {Waiting} for {Cues}: {Information} {Asymmetries} and {Challenges} in {Sharing} {Patient}-{Generated} {Data} in the {Clinic}}.
\newblock \bibinfo{journal}{\emph{Proceedings of the ACM on Human-Computer Interaction}} \bibinfo{volume}{6}, \bibinfo{number}{CSCW1} (\bibinfo{date}{March} \bibinfo{year}{2022}), \bibinfo{pages}{1--23}.
\newblock
\showISSN{2573-0142}
\href{https://doi.org/10.1145/3512954}{doi:\nolinkurl{10.1145/3512954}}


\bibitem[Pantzar and Ruckenstein(2017)]%
        {pantzar_living_2017}
\bibfield{author}{\bibinfo{person}{Mika Pantzar} {and} \bibinfo{person}{Minna Ruckenstein}.} \bibinfo{year}{2017}\natexlab{}.
\newblock \showarticletitle{Living the metrics: {Self}-tracking and situated objectivity}.
\newblock \bibinfo{journal}{\emph{DIGITAL HEALTH}}  \bibinfo{volume}{3} (\bibinfo{date}{Jan.} \bibinfo{year}{2017}).
\newblock
\showISSN{2055-2076, 2055-2076}
\urldef\tempurl%
\url{http://journals.sagepub.com/doi/10.1177/2055207617712590}
\showURL{%
\tempurl}


\bibitem[Paudel et~al\mbox{.}(2025)]%
        {paudel2025hype}
\bibfield{author}{\bibinfo{person}{Shreyasha Paudel}, \bibinfo{person}{Sabine Loos}, {and} \bibinfo{person}{Robert Soden}.} \bibinfo{year}{2025}\natexlab{}.
\newblock \showarticletitle{Hype versus Historical Continuity: Situating the Rise of AI in Climate and Disaster Risk Modeling}. In \bibinfo{booktitle}{\emph{Proceedings of the 2025 CHI Conference on Human Factors in Computing Systems}}. \bibinfo{pages}{1--17}.
\newblock


\bibitem[Project({[n.\,d.]})]%
        {WarriorCareNetwork}
\bibfield{author}{\bibinfo{person}{Wounded~Warrior Project}.} \bibinfo{year}{[n.\,d.]}\natexlab{}.
\newblock \bibinfo{title}{Warrior Care Network}.
\newblock
\urldef\tempurl%
\url{https://www.woundedwarriorproject.org/programs/warrior-care-network}
\showURL{%
Retrieved 2025-11-20 from \tempurl}


\bibitem[Qiu et~al\mbox{.}(2025)]%
        {qiu2025voice}
\bibfield{author}{\bibinfo{person}{Ling Qiu}, \bibinfo{person}{Ita~Daryanti Saragih}, \bibinfo{person}{Donna~Marie Fick}, \bibinfo{person}{S~Shyam Sundar}, {and} \bibinfo{person}{Saeed Abdullah}.} \bibinfo{year}{2025}\natexlab{}.
\newblock \showarticletitle{Voice Assistants to Deliver Cognitive Stimulation Therapy for Persons Living with Dementia}. In \bibinfo{booktitle}{\emph{Proceedings of the Extended Abstracts of the CHI Conference on Human Factors in Computing Systems}}. \bibinfo{pages}{1--7}.
\newblock


\bibitem[Rauch et~al\mbox{.}(2017)]%
        {rauch_expanding_2017}
\bibfield{author}{\bibinfo{person}{Sheila~AM Rauch}, \bibinfo{person}{Jeffrey Cigrang}, \bibinfo{person}{David Austern}, \bibinfo{person}{Ashley Evans}, {and} \bibinfo{person}{STRONG~STAR Consortium}.} \bibinfo{year}{2017}\natexlab{}.
\newblock \showarticletitle{Expanding the {Reach} of {Effective} {PTSD} {Treatment} {Into} {Primary} {Care}: {Prolonged} {Exposure} for {Primary} {Care}}.
\newblock \bibinfo{journal}{\emph{FOCUS}} \bibinfo{volume}{15}, \bibinfo{number}{4} (\bibinfo{year}{2017}), \bibinfo{pages}{406--410}.
\newblock
\showISSN{1541-4094, 1541-4108}
\href{https://doi.org/10.1176/appi.focus.20170021}{doi:\nolinkurl{10.1176/appi.focus.20170021}}


\bibitem[Rauch et~al\mbox{.}(2020)]%
        {rauch2020prolonged}
\bibfield{author}{\bibinfo{person}{Sheila~AM Rauch}, \bibinfo{person}{Barbara~Olasov Rothbaum}, \bibinfo{person}{Erin~R Smith}, {and} \bibinfo{person}{Edna~B Foa}.} \bibinfo{year}{2020}\natexlab{}.
\newblock \bibinfo{booktitle}{\emph{Prolonged exposure for PTSD in intensive outpatient programs (PE-IOP): Therapist guide}}.
\newblock \bibinfo{publisher}{Oxford University Press}.
\newblock


\bibitem[Rauch et~al\mbox{.}(2023)]%
        {rauch2023treatment}
\bibfield{author}{\bibinfo{person}{Sheila~AM Rauch}, \bibinfo{person}{Margaret~R Venners}, \bibinfo{person}{Carly Ragin}, \bibinfo{person}{Gretchen Ruhe}, \bibinfo{person}{Kristen~E Lamp}, \bibinfo{person}{Mark Burton}, \bibinfo{person}{Andrew Pomerantz}, \bibinfo{person}{Nancy Bernardy}, \bibinfo{person}{Paula~P Schnurr}, \bibinfo{person}{Jessica~L Hamblen}, {et~al\mbox{.}}} \bibinfo{year}{2023}\natexlab{}.
\newblock \showarticletitle{Treatment of posttraumatic stress disorder with prolonged exposure for primary care (PE-PC): Effectiveness and patient and therapist factors related to symptom change and retention.}
\newblock \bibinfo{journal}{\emph{Psychological Services}} (\bibinfo{year}{2023}).
\newblock


\bibitem[Rauch et~al\mbox{.}(2021)]%
        {rauch2021intensive}
\bibfield{author}{\bibinfo{person}{Sheila~AM Rauch}, \bibinfo{person}{Carly~W Yasinski}, \bibinfo{person}{Loren~M Post}, \bibinfo{person}{Tanja Jovanovic}, \bibinfo{person}{Seth Norrholm}, \bibinfo{person}{Andrew~M Sherrill}, \bibinfo{person}{Vasiliki Michopoulos}, \bibinfo{person}{Jessica~L Maples-Keller}, \bibinfo{person}{Kathryn Black}, \bibinfo{person}{Liza Zwiebach}, {et~al\mbox{.}}} \bibinfo{year}{2021}\natexlab{}.
\newblock \showarticletitle{An intensive outpatient program with prolonged exposure for veterans with posttraumatic stress disorder: Retention, predictors, and patterns of change.}
\newblock \bibinfo{journal}{\emph{Psychological services}} \bibinfo{volume}{18}, \bibinfo{number}{4} (\bibinfo{year}{2021}), \bibinfo{pages}{606}.
\newblock


\bibitem[Reger et~al\mbox{.}(2013)]%
        {reger2013pe}
\bibfield{author}{\bibinfo{person}{Greg~M Reger}, \bibinfo{person}{Julia Hoffman}, \bibinfo{person}{David Riggs}, \bibinfo{person}{Barbara~O Rothbaum}, \bibinfo{person}{Josef Ruzek}, \bibinfo{person}{Kevin~M Holloway}, {and} \bibinfo{person}{Eric Kuhn}.} \bibinfo{year}{2013}\natexlab{}.
\newblock \showarticletitle{The “PE coach” smartphone application: An innovative approach to improving implementation, fidelity, and homework adherence during prolonged exposure.}
\newblock \bibinfo{journal}{\emph{Psychological services}} \bibinfo{volume}{10}, \bibinfo{number}{3} (\bibinfo{year}{2013}), \bibinfo{pages}{342}.
\newblock


\bibitem[Rennert and Karapanos(2013)]%
        {rennert_faceit_2013}
\bibfield{author}{\bibinfo{person}{Kara Rennert} {and} \bibinfo{person}{Evangelos Karapanos}.} \bibinfo{year}{2013}\natexlab{}.
\newblock \showarticletitle{Faceit: supporting reflection upon social anxiety events with lifelogging}. In \bibinfo{booktitle}{\emph{{CHI} '13 {Extended} {Abstracts} on {Human} {Factors} in {Computing} {Systems}}}. \bibinfo{publisher}{ACM}, \bibinfo{pages}{457--462}.
\newblock
\showISBNx{978-1-4503-1952-2}
\href{https://doi.org/10.1145/2468356.2468437}{doi:\nolinkurl{10.1145/2468356.2468437}}


\bibitem[Rogers(2012)]%
        {rogers2012hci}
\bibfield{author}{\bibinfo{person}{Yvonne Rogers}.} \bibinfo{year}{2012}\natexlab{}.
\newblock \bibinfo{booktitle}{\emph{HCI theory: classical, modern, and contemporary}}. Vol.~\bibinfo{volume}{14}.
\newblock \bibinfo{publisher}{Morgan \& Claypool Publishers}.
\newblock


\bibitem[Rohani et~al\mbox{.}(2020)]%
        {rohani_mubs_2020}
\bibfield{author}{\bibinfo{person}{Darius~A Rohani}, \bibinfo{person}{Andrea Quemada~Lopategui}, \bibinfo{person}{Nanna Tuxen}, \bibinfo{person}{Maria Faurholt-Jepsen}, \bibinfo{person}{Lars~V Kessing}, {and} \bibinfo{person}{Jakob~E Bardram}.} \bibinfo{year}{2020}\natexlab{}.
\newblock \showarticletitle{{MUBS}: {A} {Personalized} {Recommender} {System} for {Behavioral} {Activation} in {Mental} {Health}}. In \bibinfo{booktitle}{\emph{Proceedings of the 2020 {CHI} {Conference} on {Human} {Factors} in {Computing} {Systems}}}. \bibinfo{publisher}{ACM}, \bibinfo{pages}{1--13}.
\newblock
\showISBNx{978-1-4503-6708-0}
\href{https://doi.org/10.1145/3313831.3376879}{doi:\nolinkurl{10.1145/3313831.3376879}}


\bibitem[Rohani et~al\mbox{.}(2019)]%
        {rohani_personalizing_2019}
\bibfield{author}{\bibinfo{person}{Darius~A Rohani}, \bibinfo{person}{Nanna Tuxen}, \bibinfo{person}{Andrea~Quemada Lopategui}, \bibinfo{person}{Maria Faurholt-Jepsen}, \bibinfo{person}{Lars~V Kessing}, {and} \bibinfo{person}{Jakob~E Bardram}.} \bibinfo{year}{2019}\natexlab{}.
\newblock \showarticletitle{Personalizing {Mental} {Health}: {A} {Feasibility} {Study} of a {Mobile} {Behavioral} {Activation} {Tool} for {Depressed} {Patients}}. In \bibinfo{booktitle}{\emph{Proceedings of the 13th {EAI} {International} {Conference} on {Pervasive} {Computing} {Technologies} for {Healthcare}}}. \bibinfo{publisher}{ACM}, \bibinfo{pages}{282--291}.
\newblock
\showISBNx{978-1-4503-6126-2}
\href{https://doi.org/10.1145/3329189.3329214}{doi:\nolinkurl{10.1145/3329189.3329214}}


\bibitem[Rose et~al\mbox{.}(2005)]%
        {rose2005counselling}
\bibfield{author}{\bibinfo{person}{Theresa Rose}, \bibinfo{person}{Del Loewenthal}, {and} \bibinfo{person}{Dennis Greenwood}.} \bibinfo{year}{2005}\natexlab{}.
\newblock \showarticletitle{Counselling and psychotherapy as a form of learning: Some implications for practice}.
\newblock \bibinfo{journal}{\emph{British Journal of Guidance \& Counselling}} \bibinfo{volume}{33}, \bibinfo{number}{4} (\bibinfo{year}{2005}), \bibinfo{pages}{441--456}.
\newblock


\bibitem[Rosen et~al\mbox{.}(2016)]%
        {rosen2016review}
\bibfield{author}{\bibinfo{person}{CS Rosen}, \bibinfo{person}{MM Matthieu}, \bibinfo{person}{S Wiltsey~Stirman}, \bibinfo{person}{JM Cook}, \bibinfo{person}{S Landes}, \bibinfo{person}{NC Bernardy}, \bibinfo{person}{KM Chard}, \bibinfo{person}{J Crowley}, \bibinfo{person}{A Eftekhari}, \bibinfo{person}{EP Finley}, {et~al\mbox{.}}} \bibinfo{year}{2016}\natexlab{}.
\newblock \showarticletitle{A review of studies on the system-wide implementation of evidence-based psychotherapies for posttraumatic stress disorder in the Veterans Health Administration}.
\newblock \bibinfo{journal}{\emph{Administration and Policy in Mental Health and Mental Health Services Research}}  \bibinfo{volume}{43} (\bibinfo{year}{2016}).
\newblock


\bibitem[Rothbaum and Schwartz(2002)]%
        {rothbaum2002exposure}
\bibfield{author}{\bibinfo{person}{Barbara~Olasov Rothbaum} {and} \bibinfo{person}{Ann~C Schwartz}.} \bibinfo{year}{2002}\natexlab{}.
\newblock \showarticletitle{Exposure therapy for posttraumatic stress disorder}.
\newblock \bibinfo{journal}{\emph{American journal of psychotherapy}} \bibinfo{volume}{56}, \bibinfo{number}{1} (\bibinfo{year}{2002}).
\newblock


\bibitem[Saka and Das(2025)]%
        {saka2025watch}
\bibfield{author}{\bibinfo{person}{Suleiman Saka} {and} \bibinfo{person}{Sanchari Das}.} \bibinfo{year}{2025}\natexlab{}.
\newblock \showarticletitle{"Watch My Health, Not My Data": Understanding Perceptions, Barriers, Emotional Impact, \& Coping Strategies Pertaining to IoT Privacy and Security in Health Monitoring for Older Adults}. In \bibinfo{booktitle}{\emph{Proceedings of the 2025 CHI Conference on Human Factors in Computing Systems}}.
\newblock


\bibitem[Sanches et~al\mbox{.}(2019)]%
        {sanches_hci_2019}
\bibfield{author}{\bibinfo{person}{Pedro Sanches}, \bibinfo{person}{Axel Janson}, \bibinfo{person}{Pavel Karpashevich}, \bibinfo{person}{Camille Nadal}, \bibinfo{person}{Chengcheng Qu}, \bibinfo{person}{Claudia Daudén~Roquet}, \bibinfo{person}{Muhammad Umair}, \bibinfo{person}{Charles Windlin}, \bibinfo{person}{Gavin Doherty}, \bibinfo{person}{Kristina Höök}, {and} \bibinfo{person}{Corina Sas}.} \bibinfo{year}{2019}\natexlab{}.
\newblock \showarticletitle{{HCI} and {Affective} {Health}: {Taking} stock of a decade of studies and charting future research directions}. In \bibinfo{booktitle}{\emph{Proceedings of the 2019 {CHI} {Conference} on {Human} {Factors} in {Computing} {Systems}}}. \bibinfo{publisher}{ACM}, \bibinfo{pages}{1--17}.
\newblock
\showISBNx{978-1-4503-5970-2}
\href{https://doi.org/10.1145/3290605.3300475}{doi:\nolinkurl{10.1145/3290605.3300475}}


\bibitem[Schertz et~al\mbox{.}(2019)]%
        {schertz_bridging_2019}
\bibfield{author}{\bibinfo{person}{Elaine Schertz}, \bibinfo{person}{Hue Watson}, \bibinfo{person}{Ashok Krishna}, \bibinfo{person}{Andrew Sherrill}, \bibinfo{person}{Hayley Evans}, {and} \bibinfo{person}{Rosa~I. Arriaga}.} \bibinfo{year}{2019}\natexlab{}.
\newblock \showarticletitle{Bridging the {Gap}: {Creating} a {Clinician}-{Facing} {Dashboard} for {PTSD}}.
\newblock In \bibinfo{booktitle}{\emph{Human-{Computer} {Interaction} – {INTERACT} 2019}}. Vol.~\bibinfo{volume}{11746}. \bibinfo{publisher}{Springer International Publishing}, \bibinfo{address}{Cham}, \bibinfo{pages}{224--233}.
\newblock
\showISBNx{978-3-030-29380-2 978-3-030-29381-9}
\href{https://doi.org/10.1007/978-3-030-29381-9_14}{doi:\nolinkurl{10.1007/978-3-030-29381-9_14}}
\newblock
\shownote{Series Title: Lecture Notes in Computer Science}.


\bibitem[Schroeder et~al\mbox{.}(2017)]%
        {schroeder_supporting_2017}
\bibfield{author}{\bibinfo{person}{Jessica Schroeder}, \bibinfo{person}{Jane Hoffswell}, \bibinfo{person}{Chia-Fang Chung}, \bibinfo{person}{James Fogarty}, \bibinfo{person}{Sean Munson}, {and} \bibinfo{person}{Jasmine Zia}.} \bibinfo{year}{2017}\natexlab{}.
\newblock \showarticletitle{Supporting {Patient}-{Provider} {Collaboration} to {Identify} {Individual} {Triggers} using {Food} and {Symptom} {Journals}}. In \bibinfo{booktitle}{\emph{Proceedings of the 2017 {ACM} {Conference} on {Computer} {Supported} {Cooperative} {Work} and {Social} {Computing}}}. \bibinfo{publisher}{ACM}, \bibinfo{pages}{1726--1739}.
\newblock
\showISBNx{978-1-4503-4335-0}
\href{https://doi.org/10.1145/2998181.2998276}{doi:\nolinkurl{10.1145/2998181.2998276}}
\newblock
\shownote{Being considered for CHR paper - how to bring PGHD into clinical practice}.


\bibitem[Schueller and Mohr(2015)]%
        {schueller_initial_2015}
\bibfield{author}{\bibinfo{person}{Stephen Schueller} {and} \bibinfo{person}{David Mohr}.} \bibinfo{year}{2015}\natexlab{}.
\newblock \showarticletitle{Initial {Field} {Trial} of a {Coach}-{Supported} {Web}-{Based} {Depression} {Treatment}}. In \bibinfo{booktitle}{\emph{Proceedings of the 9th {International} {Conference} on {Pervasive} {Computing} {Technologies} for {Healthcare}}}. \bibinfo{publisher}{ICST}.
\newblock
\showISBNx{978-1-63190-045-7}
\href{https://doi.org/10.4108/icst.pervasivehealth.2015.260115}{doi:\nolinkurl{10.4108/icst.pervasivehealth.2015.260115}}


\bibitem[Semaan et~al\mbox{.}(2017)]%
        {semaan2017military}
\bibfield{author}{\bibinfo{person}{Bryan Semaan}, \bibinfo{person}{Lauren~M Britton}, {and} \bibinfo{person}{Bryan Dosono}.} \bibinfo{year}{2017}\natexlab{}.
\newblock \showarticletitle{Military masculinity and the travails of transitioning: Disclosure in social media}. In \bibinfo{booktitle}{\emph{Proceedings of the 2017 ACM Conference on computer supported cooperative work and social computing}}. \bibinfo{pages}{387--403}.
\newblock


\bibitem[Shiner et~al\mbox{.}(2013)]%
        {shiner2013measuring}
\bibfield{author}{\bibinfo{person}{Brian Shiner}, \bibinfo{person}{Leonard~W D’Avolio}, \bibinfo{person}{Thien~M Nguyen}, \bibinfo{person}{Maha~H Zayed}, \bibinfo{person}{Yinong Young-Xu}, \bibinfo{person}{Rani~A Desai}, \bibinfo{person}{Paula~P Schnurr}, \bibinfo{person}{Louis~D Fiore}, {and} \bibinfo{person}{Bradley~V Watts}.} \bibinfo{year}{2013}\natexlab{}.
\newblock \showarticletitle{Measuring use of evidence based psychotherapy for posttraumatic stress disorder}.
\newblock \bibinfo{journal}{\emph{Administration and Policy in Mental Health and Mental Health Services Research}}  \bibinfo{volume}{40} (\bibinfo{year}{2013}), \bibinfo{pages}{311--318}.
\newblock


\bibitem[Shneiderman(2003)]%
        {shneiderman2003eyes}
\bibfield{author}{\bibinfo{person}{Ben Shneiderman}.} \bibinfo{year}{2003}\natexlab{}.
\newblock \showarticletitle{The eyes have it: A task by data type taxonomy for information visualizations}.
\newblock In \bibinfo{booktitle}{\emph{The craft of information visualization}}. \bibinfo{publisher}{Elsevier}.
\newblock


\bibitem[Sijbrandij et~al\mbox{.}(2016)]%
        {sijbrandij2016effectiveness}
\bibfield{author}{\bibinfo{person}{Marit Sijbrandij}, \bibinfo{person}{Ivo Kunovski}, {and} \bibinfo{person}{Pim Cuijpers}.} \bibinfo{year}{2016}\natexlab{}.
\newblock \showarticletitle{Effectiveness of internet-delivered cognitive behavioral therapy for posttraumatic stress disorder: A systematic review and meta-analysis}.
\newblock \bibinfo{journal}{\emph{Depression and anxiety}} \bibinfo{volume}{33}, \bibinfo{number}{9} (\bibinfo{year}{2016}).
\newblock


\bibitem[Simionato et~al\mbox{.}(2019)]%
        {simionatoBurnoutEthicalIssue2019}
\bibfield{author}{\bibinfo{person}{Gabrielle Simionato}, \bibinfo{person}{Susan Simpson}, {and} \bibinfo{person}{Corinne Reid}.} \bibinfo{year}{2019}\natexlab{}.
\newblock \showarticletitle{Burnout as an ethical issue in psychotherapy.}
\newblock \bibinfo{journal}{\emph{Psychotherapy}} \bibinfo{volume}{56}, \bibinfo{number}{4} (\bibinfo{date}{Dec.} \bibinfo{year}{2019}), \bibinfo{pages}{470--482}.
\newblock
\showISSN{1939-1536, 0033-3204}


\bibitem[Simonsen et~al\mbox{.}(2020)]%
        {simonsen2020infrastructuring}
\bibfield{author}{\bibinfo{person}{Jesper Simonsen}, \bibinfo{person}{Helena Karasti}, {and} \bibinfo{person}{Morten Hertzum}.} \bibinfo{year}{2020}\natexlab{}.
\newblock \showarticletitle{Infrastructuring and participatory design: Exploring infrastructural inversion as analytic, empirical and generative}.
\newblock \bibinfo{journal}{\emph{Computer Supported Cooperative Work (CSCW)}}  \bibinfo{volume}{29} (\bibinfo{year}{2020}).
\newblock


\bibitem[Star(1999)]%
        {star1999ethnography}
\bibfield{author}{\bibinfo{person}{Susan~Leigh Star}.} \bibinfo{year}{1999}\natexlab{}.
\newblock \showarticletitle{The ethnography of infrastructure}.
\newblock \bibinfo{journal}{\emph{American behavioral scientist}} \bibinfo{volume}{43}, \bibinfo{number}{3} (\bibinfo{year}{1999}), \bibinfo{pages}{377--391}.
\newblock


\bibitem[Star and Ruhleder(1996)]%
        {star1996steps}
\bibfield{author}{\bibinfo{person}{Susan~Leigh Star} {and} \bibinfo{person}{Karen Ruhleder}.} \bibinfo{year}{1996}\natexlab{}.
\newblock \showarticletitle{Steps toward an ecology of infrastructure: Design and access for large information spaces}.
\newblock \bibinfo{journal}{\emph{Information systems research}} \bibinfo{volume}{7}, \bibinfo{number}{1} (\bibinfo{year}{1996}), \bibinfo{pages}{111--134}.
\newblock


\bibitem[Tam et~al\mbox{.}(2023)]%
        {tam2023learning}
\bibfield{author}{\bibinfo{person}{Hannah Tam}, \bibinfo{person}{Karthik~S Bhat}, \bibinfo{person}{Priyanka Mohindra}, {and} \bibinfo{person}{Neha Kumar}.} \bibinfo{year}{2023}\natexlab{}.
\newblock \showarticletitle{Learning to Navigate Health Taboos through Online Safe Spaces}. In \bibinfo{booktitle}{\emph{Proceedings of the 2023 CHI Conference on Human Factors in Computing Systems}}. \bibinfo{pages}{1--15}.
\newblock


\bibitem[Thieme et~al\mbox{.}(2023)]%
        {thieme_designing_2023}
\bibfield{author}{\bibinfo{person}{Anja Thieme}, \bibinfo{person}{Maryann Hanratty}, \bibinfo{person}{Maria Lyons}, \bibinfo{person}{Jorge Palacios}, \bibinfo{person}{Rita~Faia Marques}, \bibinfo{person}{Cecily Morrison}, {and} \bibinfo{person}{Gavin Doherty}.} \bibinfo{year}{2023}\natexlab{}.
\newblock \showarticletitle{Designing {Human}-centered {AI} for {Mental} {Health}: {Developing} {Clinically} {Relevant} {Applications} for {Online} {CBT} {Treatment}}.
\newblock \bibinfo{journal}{\emph{ACM Transactions on Computer-Human Interaction}} \bibinfo{volume}{30}, \bibinfo{number}{2} (\bibinfo{date}{April} \bibinfo{year}{2023}).
\newblock
\showISSN{1073-0516, 1557-7325}


\bibitem[Tong et~al\mbox{.}(2025)]%
        {tong2025clinical}
\bibfield{author}{\bibinfo{person}{Fangziyun Tong}, \bibinfo{person}{Reeva Lederman}, {and} \bibinfo{person}{Simon D’Alfonso}.} \bibinfo{year}{2025}\natexlab{}.
\newblock \showarticletitle{Clinical decision support systems in mental health: A scoping review of health professionals’ experiences}.
\newblock \bibinfo{journal}{\emph{International Journal of Medical Informatics}} (\bibinfo{year}{2025}), \bibinfo{pages}{105881}.
\newblock


\bibitem[Trafton et~al\mbox{.}(2010)]%
        {trafton2010designing}
\bibfield{author}{\bibinfo{person}{Jodie~A Trafton}, \bibinfo{person}{Susana~B Martins}, \bibinfo{person}{Martha~C Michel}, \bibinfo{person}{Dan Wang}, \bibinfo{person}{Samson~W Tu}, \bibinfo{person}{David~J Clark}, \bibinfo{person}{Jan Elliott}, \bibinfo{person}{Brigit Vucic}, \bibinfo{person}{Steve Balt}, {et~al\mbox{.}}} \bibinfo{year}{2010}\natexlab{}.
\newblock \showarticletitle{Designing an automated clinical decision support system to match clinical practice guidelines for opioid therapy for chronic pain}.
\newblock \bibinfo{journal}{\emph{Implementation Science}}  \bibinfo{volume}{5} (\bibinfo{year}{2010}).
\newblock


\bibitem[Vega et~al\mbox{.}(2022)]%
        {vega_detecting_2022}
\bibfield{author}{\bibinfo{person}{Julio Vega}, \bibinfo{person}{Beth~T Bell}, \bibinfo{person}{Caitlin Taylor}, \bibinfo{person}{Jue Xie}, \bibinfo{person}{Heidi Ng}, \bibinfo{person}{Mahsa Honary}, {and} \bibinfo{person}{Roisin McNaney}.} \bibinfo{year}{2022}\natexlab{}.
\newblock \showarticletitle{Detecting {Mental} {Health} {Behaviors} {Using} {Mobile} {Interactions}: {Exploratory} {Study} {Focusing} on {Binge} {Eating}}.
\newblock \bibinfo{journal}{\emph{JMIR Mental Health}} \bibinfo{volume}{9}, \bibinfo{number}{4} (\bibinfo{year}{2022}).
\newblock
\showISSN{2368-7959}


\bibitem[Wang et~al\mbox{.}(2021)]%
        {wang_brilliant_2021}
\bibfield{author}{\bibinfo{person}{Dakuo Wang}, \bibinfo{person}{Liuping Wang}, \bibinfo{person}{Zhan Zhang}, \bibinfo{person}{Ding Wang}, \bibinfo{person}{Haiyi Zhu}, \bibinfo{person}{Yvonne Gao}, \bibinfo{person}{Xiangmin Fan}, {and} \bibinfo{person}{Feng Tian}.} \bibinfo{year}{2021}\natexlab{}.
\newblock \showarticletitle{“{Brilliant} {AI} {Doctor}” in {Rural} {Clinics}: {Challenges} in {AI}-{Powered} {Clinical} {Decision} {Support} {System} {Deployment}}. In \bibinfo{booktitle}{\emph{Proceedings of the 2021 {CHI} {Conference} on {Human} {Factors} in {Computing} {Systems}}}. \bibinfo{publisher}{ACM}, \bibinfo{pages}{1--18}.
\newblock
\showISBNx{978-1-4503-8096-6}
\href{https://doi.org/10.1145/3411764.3445432}{doi:\nolinkurl{10.1145/3411764.3445432}}


\bibitem[Wang et~al\mbox{.}(2023)]%
        {wang_human-centered_2023}
\bibfield{author}{\bibinfo{person}{Liuping Wang}, \bibinfo{person}{Zhan Zhang}, \bibinfo{person}{Dakuo Wang}, \bibinfo{person}{Weidan Cao}, \bibinfo{person}{Xiaomu Zhou}, \bibinfo{person}{Ping Zhang}, \bibinfo{person}{Jianxing Liu}, \bibinfo{person}{Xiangmin Fan}, {and} \bibinfo{person}{Feng Tian}.} \bibinfo{year}{2023}\natexlab{}.
\newblock \showarticletitle{Human-centered design and evaluation of {AI}-empowered clinical decision support systems: a systematic review}.
\newblock \bibinfo{journal}{\emph{Frontiers in Computer Science}}  \bibinfo{volume}{5} (\bibinfo{date}{June} \bibinfo{year}{2023}).
\newblock
\showISSN{2624-9898}


\bibitem[Wang et~al\mbox{.}(2016)]%
        {wang_crosscheck_2016}
\bibfield{author}{\bibinfo{person}{Rui Wang}, \bibinfo{person}{Min~SH Aung}, \bibinfo{person}{Saeed Abdullah}, \bibinfo{person}{Rachel Brian}, \bibinfo{person}{Andrew~T Campbell}, \bibinfo{person}{Tanzeem Choudhury}, \bibinfo{person}{Marta Hauser}, \bibinfo{person}{John Kane}, \bibinfo{person}{Michael Merrill}, \bibinfo{person}{Emily~A Scherer}, \bibinfo{person}{Vincent~WS Tseng}, {and} \bibinfo{person}{Dror Ben-Zeev}.} \bibinfo{year}{2016}\natexlab{}.
\newblock \showarticletitle{{CrossCheck}: toward passive sensing and detection of mental health changes in people with schizophrenia}. In \bibinfo{booktitle}{\emph{Proceedings of the 2016 {ACM} {International} {Joint} {Conference} on {Pervasive} and {Ubiquitous} {Computing}}}. \bibinfo{publisher}{ACM}, \bibinfo{pages}{886--897}.
\newblock
\showISBNx{978-1-4503-4461-6}
\href{https://doi.org/10.1145/2971648.2971740}{doi:\nolinkurl{10.1145/2971648.2971740}}


\bibitem[Wang et~al\mbox{.}(2014)]%
        {wang2014studentlife}
\bibfield{author}{\bibinfo{person}{Rui Wang}, \bibinfo{person}{Fanglin Chen}, \bibinfo{person}{Zhenyu Chen}, \bibinfo{person}{Tianxing Li}, \bibinfo{person}{Gabriella Harari}, \bibinfo{person}{Stefanie Tignor}, \bibinfo{person}{Xia Zhou}, \bibinfo{person}{Dror Ben-Zeev}, {and} \bibinfo{person}{Andrew~T Campbell}.} \bibinfo{year}{2014}\natexlab{}.
\newblock \showarticletitle{StudentLife: assessing mental health, academic performance and behavioral trends of college students using smartphones}. In \bibinfo{booktitle}{\emph{Proceedings of the 2014 ACM international joint conference on pervasive and ubiquitous computing}}. \bibinfo{pages}{3--14}.
\newblock


\bibitem[Wang et~al\mbox{.}(2018)]%
        {wang_tracking_2018}
\bibfield{author}{\bibinfo{person}{Rui Wang}, \bibinfo{person}{Weichen Wang}, \bibinfo{person}{Alex daSilva}, \bibinfo{person}{Jeremy~F. Huckins}, \bibinfo{person}{William~M. Kelley}, \bibinfo{person}{Todd~F. Heatherton}, {and} \bibinfo{person}{Andrew~T. Campbell}.} \bibinfo{year}{2018}\natexlab{}.
\newblock \showarticletitle{Tracking {Depression} {Dynamics} in {College} {Students} {Using} {Mobile} {Phone} and {Wearable} {Sensing}}.
\newblock \bibinfo{journal}{\emph{Proceedings of the ACM on Interactive, Mobile, Wearable and Ubiquitous Technologies}} \bibinfo{volume}{2}, \bibinfo{number}{1} (\bibinfo{date}{March} \bibinfo{year}{2018}), \bibinfo{pages}{1--26}.
\newblock
\showISSN{2474-9567}
\href{https://doi.org/10.1145/3191775}{doi:\nolinkurl{10.1145/3191775}}


\bibitem[Watkins et~al\mbox{.}(2018)]%
        {watkins2018treating}
\bibfield{author}{\bibinfo{person}{Laura~E Watkins}, \bibinfo{person}{Kelsey~R Sprang}, {and} \bibinfo{person}{Barbara~O Rothbaum}.} \bibinfo{year}{2018}\natexlab{}.
\newblock \showarticletitle{Treating PTSD: A review of evidence-based psychotherapy interventions}.
\newblock \bibinfo{journal}{\emph{Frontiers in behavioral neuroscience}}  \bibinfo{volume}{12} (\bibinfo{year}{2018}), \bibinfo{pages}{258}.
\newblock


\bibitem[Wenzel(2017)]%
        {wenzel2017basic}
\bibfield{author}{\bibinfo{person}{Amy Wenzel}.} \bibinfo{year}{2017}\natexlab{}.
\newblock \showarticletitle{Basic strategies of cognitive behavioral therapy}.
\newblock \bibinfo{journal}{\emph{Psychiatric Clinics}} \bibinfo{volume}{40}, \bibinfo{number}{4} (\bibinfo{year}{2017}), \bibinfo{pages}{597--609}.
\newblock


\bibitem[West et~al\mbox{.}(2018)]%
        {west_common_2018}
\bibfield{author}{\bibinfo{person}{Peter West}, \bibinfo{person}{Max Van~Kleek}, \bibinfo{person}{Richard Giordano}, \bibinfo{person}{Mark~J. Weal}, {and} \bibinfo{person}{Nigel Shadbolt}.} \bibinfo{year}{2018}\natexlab{}.
\newblock \showarticletitle{Common {Barriers} to the {Use} of {Patient}-{Generated} {Data} {Across} {Clinical} {Settings}}. In \bibinfo{booktitle}{\emph{Proceedings of the 2018 {CHI} {Conference} on {Human} {Factors} in {Computing} {Systems}}}. \bibinfo{publisher}{ACM}, \bibinfo{pages}{1--13}.
\newblock
\showISBNx{978-1-4503-5620-6}
\href{https://doi.org/10.1145/3173574.3174058}{doi:\nolinkurl{10.1145/3173574.3174058}}


\bibitem[Whealin et~al\mbox{.}(2016)]%
        {Whealin2016}
\bibfield{author}{\bibinfo{person}{Julia~M Whealin}, \bibinfo{person}{Emily~C Jenchura}, \bibinfo{person}{Ava~C Wong}, {and} \bibinfo{person}{Donna~M Zulman}.} \bibinfo{year}{2016}\natexlab{}.
\newblock \showarticletitle{How veterans with post-traumatic stress disorder and comorbid health conditions utilize ehealth to manage their health care needs: a mixed-methods analysis}.
\newblock \bibinfo{journal}{\emph{Journal of medical Internet research}} \bibinfo{volume}{18}, \bibinfo{number}{10} (\bibinfo{year}{2016}).
\newblock


\bibitem[Wilson et~al\mbox{.}(2023)]%
        {wilson2023scoping}
\bibfield{author}{\bibinfo{person}{Eric Wilson}, \bibinfo{person}{Michelle Daniel}, \bibinfo{person}{Aditi Rao}, \bibinfo{person}{Dario Torre}, \bibinfo{person}{Steven Durning}, \bibinfo{person}{Clare Anderson}, \bibinfo{person}{Nicole~H Goldhaber}, \bibinfo{person}{Whitney Townsend}, {and} \bibinfo{person}{Colleen~M Seifert}.} \bibinfo{year}{2023}\natexlab{}.
\newblock \showarticletitle{A scoping review of distributed cognition in acute care clinical decision-making}.
\newblock \bibinfo{journal}{\emph{Diagnosis}} \bibinfo{volume}{10}, \bibinfo{number}{2} (\bibinfo{year}{2023}), \bibinfo{pages}{68--88}.
\newblock


\bibitem[Wilson(1996)]%
        {wilson1996manual}
\bibfield{author}{\bibinfo{person}{G~Terence Wilson}.} \bibinfo{year}{1996}\natexlab{}.
\newblock \showarticletitle{Manual-based treatments: The clinical application of research findings}.
\newblock \bibinfo{journal}{\emph{Behaviour research and therapy}} \bibinfo{volume}{34}, \bibinfo{number}{4} (\bibinfo{year}{1996}), \bibinfo{pages}{295--314}.
\newblock


\bibitem[Wilson(1998)]%
        {wilson1998manual}
\bibfield{author}{\bibinfo{person}{G~Terence Wilson}.} \bibinfo{year}{1998}\natexlab{}.
\newblock \showarticletitle{Manual-based treatment and clinical practice.}
\newblock \bibinfo{journal}{\emph{Clinical Psychology: Science and Practice}} \bibinfo{volume}{5}, \bibinfo{number}{3} (\bibinfo{year}{1998}), \bibinfo{pages}{363}.
\newblock


\bibitem[Wilson(2007)]%
        {wilson2007manual}
\bibfield{author}{\bibinfo{person}{G~Terence Wilson}.} \bibinfo{year}{2007}\natexlab{}.
\newblock \showarticletitle{Manual-based treatment: Evolution and evaluation}.
\newblock In \bibinfo{booktitle}{\emph{Psychological clinical science}}. \bibinfo{publisher}{Routledge}, \bibinfo{pages}{126--153}.
\newblock


\bibitem[Wood et~al\mbox{.}(2017)]%
        {wood2017lies}
\bibfield{author}{\bibinfo{person}{Christopher Wood}, \bibinfo{person}{Stefan Poslad}, \bibinfo{person}{Antonios Kaniadakis}, {and} \bibinfo{person}{Jennifer Gabrys}.} \bibinfo{year}{2017}\natexlab{}.
\newblock \showarticletitle{What lies above: Alternative user experiences produced through focussing attention on GNSS infrastructure}. In \bibinfo{booktitle}{\emph{Proceedings of the 2017 Conference on Designing Interactive Systems}}. \bibinfo{pages}{161--172}.
\newblock


\bibitem[Xu et~al\mbox{.}(2024)]%
        {xu2024obviously}
\bibfield{author}{\bibinfo{person}{Tian Xu}, \bibinfo{person}{Emily Jost}, \bibinfo{person}{Laurel~H Messer}, \bibinfo{person}{Paul~F Cook}, \bibinfo{person}{Gregory~P Forlenza}, \bibinfo{person}{Sriram Sankaranarayanan}, \bibinfo{person}{Casey Fiesler}, {and} \bibinfo{person}{Stephen Voida}.} \bibinfo{year}{2024}\natexlab{}.
\newblock \showarticletitle{“Obviously, Nothing's Gonna Happen in Five Minutes”: How Adolescents and Young Adults Infrastructure Resources to Learn Type 1 Diabetes Management}. In \bibinfo{booktitle}{\emph{Proceedings of the 2024 CHI Conference on Human Factors in Computing Systems}}. \bibinfo{pages}{1--16}.
\newblock


\bibitem[Yang et~al\mbox{.}(2019)]%
        {yang_unremarkable_2019}
\bibfield{author}{\bibinfo{person}{Qian Yang}, \bibinfo{person}{Aaron Steinfeld}, {and} \bibinfo{person}{John Zimmerman}.} \bibinfo{year}{2019}\natexlab{}.
\newblock \showarticletitle{Unremarkable {AI}: {Fitting} {Intelligent} {Decision} {Support} into {Critical}, {Clinical} {Decision}-{Making} {Processes}}. In \bibinfo{booktitle}{\emph{Proceedings of the 2019 {CHI} {Conference} on {Human} {Factors} in {Computing} {Systems}}}. \bibinfo{publisher}{ACM}, \bibinfo{pages}{1--11}.
\newblock
\showISBNx{978-1-4503-5970-2}


\bibitem[Yang et~al\mbox{.}(2016)]%
        {yang_investigating_2016}
\bibfield{author}{\bibinfo{person}{Qian Yang}, \bibinfo{person}{John Zimmerman}, \bibinfo{person}{Aaron Steinfeld}, \bibinfo{person}{Lisa Carey}, {and} \bibinfo{person}{James~F. Antaki}.} \bibinfo{year}{2016}\natexlab{}.
\newblock \showarticletitle{Investigating the {Heart} {Pump} {Implant} {Decision} {Process}: {Opportunities} for {Decision} {Support} {Tools} to {Help}}. In \bibinfo{booktitle}{\emph{Proceedings of the 2016 {CHI} {Conference} on {Human} {Factors} in {Computing} {Systems}}}. \bibinfo{address}{San Jose California USA}, \bibinfo{pages}{4477--4488}.
\newblock
\showISBNx{978-1-4503-3362-7}
\urldef\tempurl%
\url{https://dl.acm.org/doi/10.1145/2858036.2858373}
\showURL{%
\tempurl}


\bibitem[Yao et~al\mbox{.}(2022)]%
        {yao2022learning}
\bibfield{author}{\bibinfo{person}{Zheng Yao}, \bibinfo{person}{Haiyi Zhu}, {and} \bibinfo{person}{Robert~E Kraut}.} \bibinfo{year}{2022}\natexlab{}.
\newblock \showarticletitle{Learning to become a volunteer counselor: Lessons from a peer-to-peer mental health community}.
\newblock \bibinfo{journal}{\emph{Proceedings of the ACM on Human-Computer Interaction}} \bibinfo{volume}{6}, \bibinfo{number}{CSCW2} (\bibinfo{year}{2022}), \bibinfo{pages}{1--24}.
\newblock


\bibitem[Zhang et~al\mbox{.}(2022)]%
        {zhang2022get}
\bibfield{author}{\bibinfo{person}{Minfan Zhang}, \bibinfo{person}{Daniel Ehrmann}, \bibinfo{person}{Mjaye Mazwi}, \bibinfo{person}{Danny Eytan}, \bibinfo{person}{Marzyeh Ghassemi}, {and} \bibinfo{person}{Fanny Chevalier}.} \bibinfo{year}{2022}\natexlab{}.
\newblock \showarticletitle{Get to the point! Problem-based curated data views to augment care for critically ill patients}. In \bibinfo{booktitle}{\emph{Proceedings of the 2022 CHI Conference on Human Factors in Computing Systems}}. \bibinfo{pages}{1--13}.
\newblock


\bibitem[Zhang et~al\mbox{.}(2021)]%
        {Zhang2021}
\bibfield{author}{\bibinfo{person}{Zhan Zhang}, \bibinfo{person}{Karen Joy}, \bibinfo{person}{Pradeepti Upadhyayula}, \bibinfo{person}{Mustafa Ozkaynak}, \bibinfo{person}{Richard Harris}, {and} \bibinfo{person}{Kathleen Adelgais}.} \bibinfo{year}{2021}\natexlab{}.
\newblock \showarticletitle{Data work and decision making in emergency medical services: a distributed cognition perspective}.
\newblock \bibinfo{journal}{\emph{Proceedings of the ACM on Human-Computer Interaction}} \bibinfo{volume}{5}, \bibinfo{number}{CSCW2} (\bibinfo{year}{2021}), \bibinfo{pages}{1--32}.
\newblock


\bibitem[Zhou et~al\mbox{.}(2022)]%
        {zhou2022veteran}
\bibfield{author}{\bibinfo{person}{Jiawei Zhou}, \bibinfo{person}{Koustuv Saha}, \bibinfo{person}{Irene~Michelle Lopez~Carron}, \bibinfo{person}{Dong~Whi Yoo}, \bibinfo{person}{Catherine~R Deeter}, \bibinfo{person}{Munmun De~Choudhury}, {and} \bibinfo{person}{Rosa~I Arriaga}.} \bibinfo{year}{2022}\natexlab{}.
\newblock \showarticletitle{Veteran critical theory as a lens to understand veterans' needs and support on social media}.
\newblock \bibinfo{journal}{\emph{Proceedings of the ACM on Human-Computer Interaction}} \bibinfo{volume}{6}, \bibinfo{number}{CSCW1} (\bibinfo{year}{2022}), \bibinfo{pages}{1--28}.
\newblock


\bibitem[Zhu et~al\mbox{.}(2017)]%
        {zhu_making_2017}
\bibfield{author}{\bibinfo{person}{Haining Zhu}, \bibinfo{person}{Yuhan Luo}, {and} \bibinfo{person}{Eun~Kyoung Choe}.} \bibinfo{year}{2017}\natexlab{}.
\newblock \showarticletitle{Making {Space} for the {Quality} {Care}: {Opportunities} for {Technology} in {Cognitive} {Behavioral} {Therapy} for {Insomnia}}. In \bibinfo{booktitle}{\emph{Proceedings of the 2017 {CHI} {Conference} on {Human} {Factors} in {Computing} {Systems}}}. \bibinfo{publisher}{ACM}, \bibinfo{pages}{5773--5786}.
\newblock
\showISBNx{978-1-4503-4655-9}
\href{https://doi.org/10.1145/3025453.3025549}{doi:\nolinkurl{10.1145/3025453.3025549}}


\bibitem[Zhu et~al\mbox{.}(2025)]%
        {zhu2025designing}
\bibfield{author}{\bibinfo{person}{Xiuqi~Tommy Zhu}, \bibinfo{person}{Heidi Cheerman}, \bibinfo{person}{Minxin Cheng}, \bibinfo{person}{Sheri~R Kiami}, \bibinfo{person}{Leanne Chukoskie}, {and} \bibinfo{person}{Eileen McGivney}.} \bibinfo{year}{2025}\natexlab{}.
\newblock \showarticletitle{Designing VR Simulation System for Clinical Communication Training with LLMs-Based Embodied Conversational Agents}. In \bibinfo{booktitle}{\emph{Proceedings of the Extended Abstracts of the CHI Conference on Human Factors in Computing Systems}}. \bibinfo{pages}{1--9}.
\newblock


\bibitem[Zwiebach et~al\mbox{.}(2019)]%
        {zwiebach2019military}
\bibfield{author}{\bibinfo{person}{Liza Zwiebach}, \bibinfo{person}{Brittany~K Lannert}, \bibinfo{person}{Andrew~M Sherrill}, \bibinfo{person}{Lauren~B McSweeney}, \bibinfo{person}{Kelsey Sprang}, \bibinfo{person}{Jessica~RM Goodnight}, \bibinfo{person}{Shaun~C Lewis}, {and} \bibinfo{person}{Sheila~AM Rauch}.} \bibinfo{year}{2019}\natexlab{}.
\newblock \showarticletitle{Military cultural competence in the context of cognitive behavioural therapy}.
\newblock \bibinfo{journal}{\emph{The Cognitive Behaviour Therapist}}  \bibinfo{volume}{12} (\bibinfo{year}{2019}).
\newblock


\end{thebibliography}


\end{document}